\documentclass[12pt]{article}

\usepackage{tabularx,ragged2e,booktabs,caption}
\usepackage{diagbox}
\usepackage[toc,page]{appendix}
\usepackage{multirow}
\usepackage{amssymb}
\usepackage{amsmath} 
\usepackage{chngpage} 
\usepackage{framed}
\usepackage{indentfirst} 
\usepackage{cite}
\usepackage{fancyhdr}  
\usepackage[authoryear]{natbib} 
\usepackage[hidelinks]{hyperref}
\usepackage[includeheadfoot,left=1.in,right=1.in,top=1.in,bottom=1.in,bindingoffset=0.in,nohead]{geometry}
\usepackage{mathtools}
\usepackage{graphicx}
\usepackage{subfigure}
\usepackage{epstopdf}
\usepackage{rotating}
\usepackage{soul}
\usepackage{lscape}

\usepackage{float}
\usepackage{enumerate}
\usepackage[onehalfspacing]{setspace}
\usepackage{url}
\usepackage{floatpag}
\floatpagestyle{empty}

\usepackage{tikz}
\usetikzlibrary{decorations, decorations.pathreplacing}
\usetikzlibrary{arrows.meta, shapes.geometric, calc, shadows}
\usepackage[draft]{tikzpeople}

\usetikzlibrary{matrix, fit}
\usetikzlibrary{backgrounds}
  
\usepackage{tcolorbox}

\usepackage{pdflscape}

\graphicspath{ {./Graphics/} }

\allowdisplaybreaks

\title{Clustered Network Connectedness: \\A New Measurement Framework\\with Application to Global Equity Markets}

\author{Bastien Buchwalter 	\vspace{-1.2mm} \\ 
	SKEMA Business School \vspace{-1.2mm}  \\ and Université Côte d'Azur \and Francis X. Diebold \vspace{-1.2mm}  \\ University of  Pennsylvania  \vspace{-1.2mm} \\ and NBER \vspace{2mm} \and Kamil Yilmaz \vspace{-1.2mm}  \\ Ko\c{c} University   \vspace{5mm} }

\date{\normalsize First Draft: July 2024\\ \vspace{-1mm} This Draft: \today}

\begin{document}

\maketitle
 
 \vspace{3mm}
 
\footnotesize

\begin{spacing}{1}  
 	
\noindent \textbf{Abstract}: Network connections, both across and within markets, are central in countless economic contexts. In recent decades, a large literature has developed and applied flexible methods for measuring network connectedness and its evolution, based on variance decompositions from vector autoregressions (VARs), as in \cite{diebold2014network}. Those VARs are, however, typically identified using full orthogonalization  \citep{Sims1980}, or no orthogonalization \citep*{koop1996impulse,PesaranShin98}, which, although useful,  are  special and extreme  cases of a more general framework that we develop in this paper.  In particular, we  allow network nodes to be connected in ``clusters", such as asset classes, industries, regions, etc., where shocks are  orthogonal across clusters (Sims style orthogonalized identification) but correlated within clusters (Koop-Pesaran-Potter-Shin style generalized identification), so that the ordering of network nodes is relevant across clusters but irrelevant within clusters. After developing the clustered connectedness framework, we apply it in a  detailed empirical exploration of sixteen country equity markets spanning three global regions.

\vspace{6mm} 
 
\noindent \textbf{Acknowledgments:} We gratefully acknowledge useful comments from the Co-Editor (Viktor Todorov), the Guest Editor, and two referees, as well as Peter Hansen, Oscar Jorda, Chang-Jin Kim, Mikkel Plagborg-Moller, and Neil Shephard.  All remaining errors are ours alone.  Yilmaz gratefully acknowledges the support of the Scientific and Technological Research Council of Turkey (TUBITAK) under grant number 121C271. 
 
\vspace{3mm}


\vspace{3mm}


\bigskip

\noindent \textbf{Contact}: bastien.buchwalter@skema.edu, fdiebold@sas.upenn.edu, kyilmaz@ku.edu.tr

\end{spacing}

\thispagestyle{empty}

\normalsize
%
%\clearpage
%{\thispagestyle{empty} \setcounter{tocdepth}{3}}
%\tableofcontents

\clearpage
\setcounter{page}{1}
\thispagestyle{empty}

\section{Introduction} \label{intro}

\setcounter{page}{1}
     
\textit{Network connectedness}  is of interest in numerous economic contexts, from financial markets, to business cycles, to international trade.  In this paper we propose a flexible approach to empirical network connectedness measurement and interpretation, working in the Diebold-Yilmaz (DY) framework \citep{diebold2009measuring}, which has been developed and applied extensively in recent decades \citep{DY2023}.

In the DY framework, one proceeds in several steps. First, one fits a dynamic approximating model to the $N$-dimensional set of objects (network nodes) $y$ whose connectedness is to be measured. Vector autoregressions (VARs) are widely used as approximating models, because of their appealing blend of approximation accuracy and simplicity. Hence we will refer to the approximating model as a ``VAR" throughout this paper, but other models, including structural models, may  be used.

Second, having estimated an approximating VAR, one uses it to produce impulse-response functions (IRFs) or variance decompositions (VDs).  DY-style empirical connectedness measurement seeks consistency with a wide variety of unknown underlying data-generating processes, and it therefore seeks to impose only minimal restrictions when identifying IRFs/VDs.

 Two approaches to IRF/VD identification, namely the ``Cholesky-factor approach" and the ``generalized approach", are dominant in the empirical DY connectedness measurement literature.\footnote{We also note the connectedness measurement potential of VDs obtained not via traditional Cholesky or generalized VAR-based methods, but rather via direct local-projection methods, as in \cite{Jorda2005} and \cite{Mikkel2021}, and surveyed for example in \cite*{JordaTaylor2024}. We leave exploration of that avenue to future research.} In the Cholesky-factor approach, the  IRFs/VDs are obtained from Cholesky-factor orthogonalizing transformations of the reduced-form VAR shocks  \citep{Sims1980}. Because shocks are orthogonal after Cholesky transformation, one can obtain the causal impact of a shock to $y_j$ on $y_i$ (the IRF object of interest), or the corresponding fraction of the optimal forecast error variance of $y_i$ due to shocks originating with $y_j$ (the VD object of interest), for any $i$ and $j$.\footnote{\cite{RambachanAndShephard2021} provide a deep analysis of the conditions under which IRFs and VDs may be given causal interpretation, and how those conditions relate to VAR identification.} But this benefit comes at a cost: the resulting IRFs/VDs can depend importantly on the variable ordering, and the number of possible orderings grows massively (indeed factorially) with the VAR dimension $N$. Hence $N$ must be very small when using Cholesky-factor IRFs/VDs if one hopes, for example, to check robustness to ordering.	
	
In the generalized approach to IRF/VD identification, the IRFs/VDs are obtained directly from the reduced-form VAR shocks, without orthogonalization \citep*{koop1996impulse, PesaranShin98}. The benefits and costs of the generalized approach are precisely opposite those of the orthogonalized approach. On the one hand, the generalized approach effectively treats each variable as if it were first in a Cholesky ordering, so that IRF/VD results do not depend on ordering, which allows for inclusion of  many variables in the VAR without worry about robustness to ordering. On the other hand, the generalized IRF/VD results capture co-movement (correlation) but not contagion (causality).

Finally, having obtained VDs, one uses network theory to draw  connectedness implications. In particular, a VD matrix may be interpreted as the adjacency matrix of a weighted, directed network \citep{diebold2014network}, so that connectedness is characterized by the VD network in- and out-degrees,  the degree distribution and its moments (particularly the mean degree), etc. 
 
In this paper we extend the DY connectedness measurement framework to simultaneously incorporate  \textit{both} orthogonalized and generalized identification, by allowing for shock clusters (or blocks, or groups, corresponding to different asset classes, industries, regions, etc.), such that shocks are orthogonal across clusters but correlated within clusters. This allows us to bridge the divide between Forbes-Rigobon (2002) ``contagion" (empirically captured by orthogonalized IRFs and VDs -- that is, $N$ ``clusters" of size 1) and ``co-movement" (empirically captured by generalized IRFs and VDs -- that is, 1  cluster of size $N$), which emerge as very special cases. Moreover, it pragmatically allows for incorporation of causal ordering while simultaneously keeping the number of possible orderings small, because ordering is relevant only across clusters, not within them.

We proceed as follows. In section \ref{two} we present orthogonalized, generalized, and clustered VAR identifications for IRFs and VDs. The IRF perspective is more effective for introducing the relevant concepts and  issues, so we begin with it in section \ref{back} in orthogonalized and generalized identification contexts. However,  the VD perspective is more effective for actual connectedness measurement, so we move to it in section \ref{clus}, where we introduce specific connectedness measures in clustered contexts. In section  \ref{stocks} we maintain the VD connectedness perspective, illustrating and comparing various connectedness measures, and comparing results from clustered vs  generalized identifications, in a detailed empirical exploration of equity markets for sixteen countries spanning three global regions. We conclude in section \ref{concl}.

\section{Measuring Network Connectedness}  \label{two}

\subsection{Background} \label{back}
    
In this section we introduce our network framework in several steps.  We will refer to network nodes as ``assets" or ``asset returns". This serves two purposes. First, it lends economy and concreteness to the discussion, just as with our use of ``VARs" rather than ``network approximating models". And second, it sets the stage for our subsequent exploration of global equity markets in section \ref{stocks}. We emphasize, however, that our application to global equities is just one example  (albeit a very important example), and that even within the global equity space our framework could be applied not only to returns, but also to return volatilities, liquidities, etc.\footnote{As an example of a different application area, consider product prices within vs across industry sectors. One can easily imagine a scenario where structural shocks to within-sector prices are correlated, but cross-sector prices are approximately orthogonal.}

\subsubsection{Basic VAR Framework}
  
Consider a covariance-stationary \(N\)-variable $P^{\rm th}$-order VAR,
\begin{equation} 
\boldsymbol{x}_t=  \sum_{p=1}^P\boldsymbol{\Phi}_p \boldsymbol{x}_{t-p} + \boldsymbol{u}_t = \sum_{i=0}^\infty \boldsymbol{A}_i\boldsymbol{u}_{t-i}, \hspace{22pt}t=1,2,\dots, T, \label{VAR}
\end{equation}
where  \(\boldsymbol{x}_t = \lbrack  x_{1t} \ x_{2t} \ \dots  x_{Nt} \ \rbrack'\)
is an \([N\times 1]\) vector of asset returns and  \(\boldsymbol{\Phi}_p\) is an \([N \times N]\) parameter matrix for lag \(p\). Further, \(\mathbb{E}[\boldsymbol{u}_t]=0\) and \(\mathbb{V} \lbrack \boldsymbol{u}_t \rbrack=\boldsymbol{\Sigma}\) for all \(t\), where \(\boldsymbol{\Sigma}=\{ \sigma_{ij}, i,j=1,2, \dots N \} \) is an \([N \times N]\) symmetric positive semi-definite matrix.  Finally, \(\boldsymbol{A}_0\) is an  \([N\times N]\)  identity matrix, and \(\boldsymbol{A}_i=\boldsymbol{\Phi}_1 \boldsymbol{A}_{i-1} +\boldsymbol{\Phi}_2 \boldsymbol{A}_{i-2} +\dots+\boldsymbol{\Phi}_p \boldsymbol{A}_{i-p}  \), $i = 1, 2, ...$ (with \(\boldsymbol{A}_i\) an \([N\times N]\) zero matrix for  \(i<0\)). For lower-triangular non-singular $[N \times N]$ matrix  $\boldsymbol{Q}_C$,  we can rewrite the moving average representation in \eqref{VAR} without loss of generality as
\begin{equation}
\boldsymbol{x}_t= \sum_{i=0}^\infty \boldsymbol{A}_i\boldsymbol{u}_{t-i} 
= \sum_{i=0}^\infty (\boldsymbol{A}_i\boldsymbol{Q}_C) (\boldsymbol{Q}_C^{-1}\boldsymbol{u}_{t-i})  \nonumber
= \sum_{i=0}^\infty ( \boldsymbol{A}_i\boldsymbol{Q}_C ) \, \boldsymbol{\epsilon}_{t-i}, \nonumber
\end{equation}
where \(\mathbb{E}[\boldsymbol{\epsilon}_{t}]=\mathbb{E}[\boldsymbol{Q}_C^{-1}\boldsymbol{u}_t]=0\), and    $\mathbb{V}[\boldsymbol{\epsilon}_{t}]=\mathbb{V}[\boldsymbol{Q}_C^{-1}\boldsymbol{u}_t] \) =  \(\boldsymbol{\Omega}_C$ with elements $\{ \omega_{C,ij}, i,j=1,2, \dots N \}$.

Assuming linearity of conditional expectations as in  \cite*{koop1996impulse}, the $N$-vector of $h$-step responses of the elements of $\boldsymbol{x}$ to a $\delta_j$ shock in $\epsilon_{jt}$ is 
$$
\boldsymbol{\psi}_j^C(h) = ( \boldsymbol{A}_h \boldsymbol{Q}_C) \, \mathbb{E} [ \boldsymbol{\epsilon}_t  \lvert  \epsilon_{j,t}  =  \delta_j ]
= \frac{ ( \boldsymbol{A}_h \boldsymbol{Q}_C)   \, \boldsymbol{\Omega}_C \boldsymbol{e}_j  \delta_j}{\omega_{C,jj}},
$$
 where \(\boldsymbol{e}_j\) is a  \([N \times 1]\) selection vector with one in the \(j^{th}\) position and zero elsewhere.  Setting $\delta_j = \sqrt{\omega_{C,jj}}$ gives the responses to a one standard deviation shock in  $\epsilon_{jt}$,
\begin{equation}
\boldsymbol{\psi}_j^C(h) = \frac{( \boldsymbol{A}_h \boldsymbol{Q}_C) \, \boldsymbol{\Omega}_C \boldsymbol{e}_j}{\sqrt{\omega_{C,jj}}}. \label{scaled2}
\end{equation}

In concluding this section, let us say a bit more about the nonsingular matrix \(\boldsymbol{Q}_C\), which  plays a crucial role in what follows but has not yet been discussed. For now, suffice it to say that \(C\) will denote the number of clusters, which determines the structure of \(\boldsymbol{Q}_C\) and hence $\boldsymbol{\Omega}_C$, both of which impact the  impulse response \eqref{scaled2}. While any number of clusters \(C \in [1, N]\) could be operative, the literature has so far only focused on two very special  cases: \(C=N\), where each network node  is its own cluster, and \(C=1\), where all nodes are grouped into a single cluster. These two settings translate respectively to orthogonalized and generalized impulse response functions, to which we now turn.

\subsubsection{Orthogonalized Impulse Responses $(C=N)$}

As the name indicates, in this approach the structural VAR shocks are uncorrelated. This means that each asset represents a cluster and there is no correlation across  clusters, i.e. there are as many clusters, \(C\), as there are assets, $N$. Mathematically, this is achieved by setting $C=N$, which translates into $\boldsymbol{Q}_N=\boldsymbol{M}$,  where $\boldsymbol{M}$ is the unique lower-triangular matrix that satisfies the Cholesky decomposition,  $\boldsymbol{MM}'=\boldsymbol{\Sigma}$. Further, in this approach the variance-covariance matrix of the underlying structural VAR shocks, $\boldsymbol{\Omega}_N=\mathbb{V}[\boldsymbol{Q}_N^{-1}\boldsymbol{u}_{t}]=\mathbb{V}[\boldsymbol{M}^{-1}\boldsymbol{u}_{t}]$, is an [$N \times N$] identity matrix $\boldsymbol{I}_N$, so that $\sqrt{\omega_{N,jj}}=1$ and $\sqrt{\omega_{N,ij}}=0$ for $i\neq j$.  This yields the well known orthogonalized impulse response functions introduced by \cite{Sims1980},
\begin{align}
\boldsymbol{\psi}^N_j(h)=\frac{\boldsymbol{A}_h \boldsymbol{Q}_N\boldsymbol{\Omega}_N \boldsymbol{e}_j}{\sqrt{\omega_{N,jj}}} =\frac{\boldsymbol{A}_h \boldsymbol{M}\boldsymbol{I}_N \boldsymbol{e}_j}{\sqrt{1}} 
=\boldsymbol{A}_h\boldsymbol{Me}_j=\boldsymbol{\psi}^o_j(h), \label{ort}
\end{align}
where the superscript  $o$ indicates ``orthogonalized" and it is understood that $C=N$.

\subsubsection{Generalized Impulse Responses $(C=1)$}

Generalized impulse responses go to the opposite extreme. That is, rather than putting each asset in its own cluster and removing all correlation across shocks, as in the orthogonalized approach, generalized impulse responses allow for correlated shocks. In particular, the generalized approach uses the reduced-form VAR shocks depicted in equation (\ref{VAR}). Intuitively, this translates to the idea of having only one cluster, i.e. setting $C=1$, and allowing for correlated shocks within that single cluster. From a mathematical perspective, this translates to $\boldsymbol{Q_1} = \boldsymbol{I}_N$. It follows that $\boldsymbol{\Omega}_1=\mathbb{V}[\boldsymbol{Q}_1^{-1}\boldsymbol{u}_{t-i}]=\mathbb{V}[\boldsymbol{I}_N^{-1}\boldsymbol{u}_{t-i}]=\mathbb{V}[\boldsymbol{u}_{t-i}]=\boldsymbol{\Sigma}$ and that $\omega_{1,jj}=\sigma_{jj},~ \forall j$. This yields the generalized impulse responses introduced by \cite*{koop1996impulse},
\begin{align}
\boldsymbol{\psi}^1_j(h)
=\frac{\boldsymbol{A}_h \boldsymbol{Q}_1\boldsymbol{\Omega}_1 \boldsymbol{e}_j}{\sqrt{\omega_{1,jj}}}
=\frac{\boldsymbol{A}_h \boldsymbol{I}_N\boldsymbol{\Sigma} \boldsymbol{e}_j}{\sqrt{\omega_{1,jj}}}
=\frac{\boldsymbol{A}_h\boldsymbol{\Sigma e}_j}{\sqrt{\sigma_{jj}}}=\boldsymbol{\psi}^g_j(h), \label{gen}
\end{align} 
where the superscript $g$ indicates ``generalized"  and it is understood that $C=1$.

\subsubsection{Discussion} 

The benefits  of  orthogonalized IRFs stem from their ability to quantify the causal impacts of shocks. But those benefits come at a cost: The outcome can depend crucially on the variable ordering, and the number of possible orderings grows factorially with the number of assets, which prohibits checking IRF robustness to ordering except in very low-dimensional VARs. The benefits and costs of generalized IRFs are precisely opposite. The generalized approach avoids the issue of ordering, but that benefit comes at the cost of quantifying only co-movement, not  contagion.
 
Against this background, in what follows we build on the seminal work of  \cite{forbes2002no}, who ask whether asset-return connections are better characterized as co-movement or contagion (``spillovers").  We progress by effectively allowing for \textit{both} co-movement and contagion, integrating the  orthogonalized and generalized IRF approaches via \textit{clustering}, to which we now turn.
 
% \clearpage
   
\subsection{Clustering}  \label{clus}

\subsubsection{Cluster-Orthogonalized Impulse Responses  $(C \in [1, N])$}
 
Consider the reduced-form VAR given by equation \eqref{VAR}, and suppose that its variables can be grouped into $C$ known clusters with similar characteristics, such as asset classes (e.g., stocks, bonds, commodities, etc.) or regions (e.g., North America, Europe, East Asia, etc.).  When  $C=3$, for example, we might have  reduced-form VAR shock covariance matrix
\begin{equation}
	\scriptsize
\begin{tikzpicture}[baseline=(current  bounding  box.center)]
	\node[text width=2cm] at (-4.5,0)  {\large{$\boldsymbol{\Sigma}_3 ~ $}};
	\node[text width=2cm] at (-3.75,0) { \scriptsize{=}}; \matrix[
	matrix of math nodes,
	row sep=2.2ex,
	column sep=2.0ex,
	inner sep=2pt,
	outer sep=0pt,
	left delimiter=( , right delimiter = ) ,     
	nodes={text width=1.8em, text height=1.45ex, text depth=1.6ex, align=center}
	] (m) 
	{
		\sigma_{11} &  \sigma_{12} &  \sigma_{13} &  \sigma_{14} &\sigma_{15} &  \sigma_{16} &  \sigma_{17} &  \dots & \sigma_{1N} \\
		\sigma_{21} &  \sigma_{22} &  \sigma_{23} &  \sigma_{24} &\sigma_{25} &  \sigma_{26} &  \sigma_{27} &  \dots & \sigma_{2N} \\
		\sigma_{31} &  \sigma_{32} &  \sigma_{33} &  \sigma_{34} &\sigma_{35} &  \sigma_{36} &  \sigma_{37} &  \dots & \sigma_{3N} \\
		\sigma_{41} &  \sigma_{42} &  \sigma_{43} &  \sigma_{44} &\sigma_{45} &  \sigma_{46} &  \sigma_{47} &  \dots & \sigma_{4N} \\
		\sigma_{51} &  \sigma_{52} &  \sigma_{53} &  \sigma_{54} &\sigma_{55} &  \sigma_{56} &  \sigma_{57} &  \dots & \sigma_{5N} \\
		\sigma_{61} &  \sigma_{62} &  \sigma_{63} &  \sigma_{64} &\sigma_{65} &  \sigma_{66} &  \sigma_{67} &  \dots & \sigma_{6N} \\
		\sigma_{71} &  \sigma_{72} &  \sigma_{73} &  \sigma_{74} &\sigma_{75} &  \sigma_{76} &  \sigma_{77} &  \dots & \sigma_{7N} \\
		\vdots & \vdots & \vdots & \vdots & \vdots & \vdots & \vdots &\ddots &\vdots  \\
		\sigma_{N1} &  \sigma_{N2} &  \sigma_{N3} &  \sigma_{N4} &\sigma_{N5} &  \sigma_{N6} &  \sigma_{N7} & \dots & \sigma_{NN} \\
	};
	\begin{scope}[on background layer]
		\node[fit=(m-1-1)(m-3-3), draw=black!100,  thick, dashed, fill=lightgray!0, rounded corners] {};
		\node[fit=(m-4-4)(m-5-5), draw=black!100,  thick, dashed, fill=lightgray!0, rounded corners] {};
		\node[fit=(m-6-6)(m-9-9), draw=black!100, thick, dashed, fill=lightgray!0, rounded corners] {};
	\end{scope};  
	\node[text width=2cm] at (5.4,0) {$=$};
	\begin{scope}[xshift=6.1cm]\matrix[
		matrix of math nodes,
		row sep=1.5ex,
		column sep=1.9ex,
		outer sep=-1pt,
		left delimiter=( , right delimiter = ) ,
		nodes={text width=1.3em, text height=1.75ex, text depth=.2ex, align=center}
		] (m) 
		{
			\boldsymbol{\Sigma}_{11} & \boldsymbol{ \Sigma}_{12}& \boldsymbol{ \Sigma}_{13} \\
			\boldsymbol{\Sigma}_{21} & \boldsymbol{ \Sigma}_{22}&  \boldsymbol{\Sigma}_{23} \\
			\boldsymbol{\Sigma}_{31} & \boldsymbol{ \Sigma}_{32}&  \boldsymbol{\Sigma}_{33} \\
		} ;\end{scope};
	\begin{scope}[on background layer]
		\node[fit=(m-1-1)(m-1-1), draw=black!100, thick, dashed, fill=lightgray!0, rounded corners, xshift=0.5mm] {};
		\node[fit=(m-2-2)(m-2-2), draw=black!100,  thick, dashed, fill=lightgray!0, rounded corners, xshift=0.5mm] {};
		\node[fit=(m-3-3)(m-3-3), draw=black!100, thick, dashed, fill=lightgray!0, rounded corners, xshift=0.5mm] {};
	\end{scope}  
\end{tikzpicture}
	\normalsize \label{Generalized}
\end{equation} 

\vspace{3mm} 

\noindent where the clustering is indicated by dashed boxes.\footnote{We use $C=3$ clusters purely for illustrative concreteness;  there is no loss of generality, as the framework that we will soon discuss holds for any number \(C\) of potential clusters (as long as \(C\leq N\)) and with any number of assets \(N_c\) in a given cluster \(c\) (as long as \(\sum_{c=1}^C N_c=N\)).} 
  
Note that we are not assuming block diagonality of $\boldsymbol{\Sigma}$, as its off-diagonal blocks are generally non-zero, as in equation \eqref{Generalized} immediately above.  Rather, we are effectively assuming block diagonality of the corresponding underlying \textit{structural} shock covariance matrix $\boldsymbol{\Omega}$, which neither  generalized nor orthogonalized impulse responses can accommodate, because they both place overly-restrictive structure on \(\boldsymbol{Q}_C\). As we have seen, generalized IRFs consider all assets to be part of a single cluster (\(C=1\)) and set \(\boldsymbol{Q}_1=\boldsymbol{I}_N\), which forces the underlying structural shock covariance matrix to be  \(\boldsymbol{\Omega}_1=\boldsymbol{\Sigma}\).  Alternatively,  orthogonalized IRFs consider each asset to represent a single uncorrelated cluster (\(C=N\)) and set  \(\boldsymbol{Q}_N=\boldsymbol{M}\), which forces the underlying structural shock covariance matrix to be  \(\boldsymbol{\Omega}_N=\boldsymbol{I}_N\). 

What is needed is a $\boldsymbol{Q}$ matrix consistent with block-diagonality of the underlying structural shock covariance matrix $\boldsymbol{\Omega}$; that is, a $\boldsymbol{Q}$ matrix such that $\boldsymbol{Q}^{-1}$  orthogonalizes reduced-form residuals across, but not within, clusters. Returning to our $C=3$ example, we need a $\boldsymbol{Q}_3$ such that $\boldsymbol{Q}_3^{-1} \boldsymbol{u_t}$ has covariance matrix
\begin{equation}
\scriptsize
\begin{tikzpicture}[baseline=(current  bounding  box.center)]
	\node[text width=2cm] at (-4.6,-.0272)  {\large{$ \boldsymbol{\Omega}_3$}};
	\node[text width=2cm] at (-3.9,0) { \scriptsize{=}}; \matrix[
	matrix of math nodes,
	row sep=2.2ex,
	column sep=2.0ex,
	inner sep=2pt,
	outer sep=0pt,
	left delimiter=( , right delimiter = ) ,
	nodes={text width=1.95em, text height=1.45ex, text depth=1.6ex, align=center}
	] (m)
	{
		\omega_{3,11} &  \omega_{3,12} & \omega_{3,13}  & 0 & 0 & 0 & 0 &  \dots & 0 \\
		\omega_{3,21} &  \omega_{3,22} & \omega_{3,23}  & 0 & 0 & 0 & 0 &  \dots & 0 \\
		\omega_{3,31}  & \omega_{3,32}  & \omega_{3,33} & 0 & 0 & 0 & 0 &  \dots & 0 \\
		0 & 0 & 0 &  \omega_{3,44} &\omega_{3,45} & 0& 0 &  \dots & 0 \\
		0& 0& 0&  \omega_{3,54} &\omega_{3,55} &  0 & 0 &  \dots & 0 \\
		0& 0&  0 &  0 &0 &  \omega_{3,66} & \omega_{3,67}&  \dots & \omega_{3,6N} \\
		0 & 0 & 0 & 0 & 0 & \omega_{3,76} & \omega_{3,77}  & \dots & \omega_{3,7N} \\
		\vdots & \vdots & \vdots & \vdots & \vdots & \vdots & \vdots &\ddots &\vdots  \\
		0 & 0 & 0 & 0 & 0 & \omega_{3,N6} &\omega_{3,N7}& \dots &\hspace{-4pt} \omega_{3,NN} \\
	};
	\begin{scope}[on background layer]
		\node[fit=(m-1-1)(m-3-3), draw=black!100,  thick, dashed, fill=lightgray!0, rounded corners, xshift=0.5mm] {};
		\node[fit=(m-4-4)(m-5-5), draw=black!100,  thick, dashed, fill=lightgray!0, rounded corners, xshift=0.5mm] {};
		\node[fit=(m-6-6)(m-9-9), draw=black!100, thick, dashed, fill=lightgray!0, rounded corners, xshift=0.9mm] {};
	\end{scope};
	\node[text width=2cm] at (5.65,0) {$=$};
	\begin{scope}[xshift=6.6cm]\matrix[
		matrix of math nodes,
		row sep=1.6ex,
		column sep=1.9ex,
		outer sep=-.5pt,
		left delimiter=( , right delimiter = ) ,
		nodes={text width=1.7em, text height=1.75ex, text depth=.02ex, align=center}
		] (m)
		{
			\boldsymbol{\Omega}_{3,11} &  \textbf{0}&   \textbf{0} \\
			\textbf{0} &  \boldsymbol{\Omega}_{3,22} &   \textbf{0}\\
			\textbf{0} &   \textbf{0} &  \boldsymbol{\Omega}_{3,33}  \\
		} ;\end{scope};
	\begin{scope}[on background layer]
		\node[fit=(m-1-1)(m-1-1), draw=black!100, thick, dashed, fill=lightgray!0, rounded corners, xshift=1mm] {};
		\node[fit=(m-2-2)(m-2-2), draw=black!100,  thick, dashed, fill=lightgray!0, rounded corners, xshift=1mm] {};
		\node[fit=(m-3-3)(m-3-3), draw=black!100, thick, dashed, fill=lightgray!0, rounded corners, xshift=1mm] {};
	\end{scope}
\end{tikzpicture}.
\normalsize \label{Sectoral}
\end{equation}

\vspace{3mm}

\noindent In Appendix \ref{Q3} we show by sequential linear projection that the relevant $\boldsymbol{Q}^{-1}$ matrix is
\begin{equation} \label{clusterize}
	\boldsymbol{Q}_3^{-1} = 
	\begin{pmatrix}
		\boldsymbol{I} &0 &0\\
		-\boldsymbol{\Sigma}_{21}\boldsymbol{\Sigma}_{11}^{-1}& \boldsymbol{I}  &0 \\
		- \boldsymbol{\Sigma}_{31}\boldsymbol{\Sigma}^{11}-\boldsymbol{\Sigma}_{32}\boldsymbol{\Sigma}^{21}&-\boldsymbol{\Sigma}_{31}\boldsymbol{\Sigma}^{12}-\boldsymbol{\Sigma}_{32}\boldsymbol{\Sigma}^{22}& \boldsymbol{I}
	\end{pmatrix},
\end{equation}
where
\begin{equation}
	\begin{pmatrix}
		\boldsymbol{\Sigma}^{11} &  \boldsymbol{\Sigma}^{12} \\
		\boldsymbol{\Sigma}^{21} &  \boldsymbol{\Sigma}^{22}
	\end{pmatrix}
	\equiv
	\begin{pmatrix}
		\boldsymbol{\Sigma}_{11} &  \boldsymbol{\Sigma}_{12} \\
		\boldsymbol{\Sigma}_{21} &  \boldsymbol{\Sigma}_{22}
	\end{pmatrix}^{-1},
\end{equation}
and we provide formulae for the  $\Sigma^{ij}$'s.

\subsubsection{Cluster-Orthogonalized Variance Decompositions}

Thus far we have focused exclusively on IRFs, where basic issues and identification concepts are most easily introduced, but forecast error variance decompositions (VDs), which are simple transformations of IRFs, turn out to be more appealing for constructing and applying actual connectedness measures. First, like IRFs, VDs make obvious intuitive sense  and answer a key connectedness question, namely (at the most granular pairwise level) ``How much of the $H$-step-ahead uncertainty in asset return $i$ is due to shocks originating  from return $j$?" Second, VDs  also easily allow for levels of cross-sectional aggregation beyond pairwise, answering broader questions like ``How much of the future uncertainty in one return is due to shocks from \textit{all other} returns?".\footnote{In contrast, aggregative connectedness measurement is trickier with IRFs, which as routinely studied have a pairwise orientation. Hence, for example, attempts at IRF aggregation must confront the fact that positive and negative responses can offset, unlike variance shares, all of which must be positive.}  Third,  VDs  easily allow not only for cross-sectional aggregation, but also for temporal aggregation, via different connectedness strengths at different horizons $H$, facilitating  examination of a variety of horizons (and selection of a  preferred horizon if desired). Finally, the matrix of  VDs can be viewed as the adjacency matrix of a weighted directed network, as emphasized in \cite{diebold2014network}, bringing powerful network perspectives and tools in touch with connectedness measurement. 

We denote the $H$-step-ahead VD by \(\tilde{\theta}^{C}_{ij}(H)\):
\begin{align}
\tilde{\theta}^{C}_{ij}(H) = \frac{\sum_{h=0}^{H-1} (\boldsymbol{e}_i' \boldsymbol{\psi}_j^C(h))^2}{\sum_{h=0}^{H-1}(\boldsymbol{e}_i' \boldsymbol{A}_h \boldsymbol{\Sigma} \boldsymbol{A}_h'  \boldsymbol{e}_i)^2} = \frac{\omega_{C,jj}^{-1}\sum_{h=0}^{H-1} (\boldsymbol{e}_i' \boldsymbol{A}_h\boldsymbol{Q}_C\boldsymbol{\Omega}_C \boldsymbol{e}_j)^2}{\sum_{h=0}^{H-1}(\boldsymbol{e}_i' \boldsymbol{A}_h \boldsymbol{\Sigma} \boldsymbol{A}_h' \boldsymbol{e}_i)^2}, \label{thetasec}
\end{align}
where \(\tilde{\theta}_{ij}^C\) is the share of the $H$-step-ahead forecast error variance of asset $i$ due to shocks from asset $j$. In parallel to the IRF equations (\ref{ort}) and (\ref{gen}), the VD equation (\ref{thetasec}) nests both orthogonalized and generalized versions:
\begin{equation*}
\tilde{\theta}^o_{ij}(H) = \tilde{\theta}^{N}_{ij}(H) = \frac{\sum_{h=0}^{H-1} (\boldsymbol{e}_i' \boldsymbol{A}_h \boldsymbol{M} \boldsymbol{e}_j)^2}{\sum_{h=0}^{H-1}(\boldsymbol{e}_i' \boldsymbol{A}_h \boldsymbol{\Sigma} \boldsymbol{A}_h' \boldsymbol{e}_i)^2}
\end{equation*}
and
\begin{equation*}
	\tilde{\theta}^g_{ij}(H) = \tilde{\theta}^{1}_{ij}(H) = \frac{\sigma_{jj}^{-1}\sum_{h=0}^{H-1} (\boldsymbol{e}_i'  \boldsymbol{A}_h\boldsymbol{\Sigma} \boldsymbol{e}_j)^2}{\sum_{h=0}^{H-1}(\boldsymbol{e}_i' \boldsymbol{A}_h \boldsymbol{\Sigma} \boldsymbol{A}_h' \boldsymbol{e}_i)^2}.
\end{equation*}

We note that \(\sum_{j=1}^N\tilde{\theta}_{ij}^o(H)=1\), while generally \(\sum_{j=1}^N\tilde{\theta}_{ij}^C(H)\neq1\), and indeed \(\sum_{j=1}^N\tilde{\theta}_{ij}^g(H)\neq1\). This is due to the non-zero covariance of residuals in the case of  clustered  and generalized shocks. However, in line with \cite{diebold2012better}, we can normalize to produce  $\theta_{ij}^C(H) = \frac{\tilde{\theta}_{ij}^C(H)}{ \sum_{j=1}^N\tilde{\theta}_{ij}^C(H)}$  
and $\theta_{ij}^g(H)=\frac{\tilde{\theta}_{ij}^g(H)}{\sum_{j=1}^N\tilde{\theta}_{ij}^g(H)}$, so that \(\sum_{j=1}^N\theta_{ij}^C(H)=1\), and \(\sum_{j=1}^N\theta_{ij}^g(H)=1\).

A graphical VD illustration that matches the three-cluster structure of $\Sigma$ in equation \eqref{Generalized} and $\Omega$ in equation \eqref{Sectoral} is

\begin{equation}
\scriptsize
\begin{tikzpicture}[baseline=(current  bounding  box.center)]
	\node[text width=2cm] at (-5.5,-.0272)  {\large{$\Theta^C(H)$}};
	\node[text width=2cm] at (-3.9,-.0272) { \scriptsize{=}}; \matrix[
	matrix of math nodes,
	row sep=2.2ex,
	column sep=2.3ex,
	inner sep=2pt,
	outer sep=-1pt,
	left delimiter=( , right delimiter = ) ,
	nodes={text width=1.85em, text height=1.45ex, text depth=1.6ex, align=center}
	] (m) 
	{
		\theta_{11} &  \theta_{12} &  \theta_{13} &  \theta_{14} &\theta_{15} &  \theta_{16} &  \theta_{17} &  \dots & \theta_{1N} \\
		\theta_{21} &  \theta_{22} &  \theta_{23} &  \theta_{24} &\theta_{25} &  \theta_{26} &  \theta_{27} &  \dots & \theta_{2N} \\
		\theta_{31} &  \theta_{32} &  \theta_{33} &  \theta_{34} &\theta_{35} &  \theta_{36} &  \theta_{37} &  \dots & \theta_{3N} \\
		\theta_{41} &  \theta_{42} &  \theta_{43} &  \theta_{44} &\theta_{45} &  \theta_{46} &  \theta_{47} &  \dots & \theta_{4N} \\
		\theta_{51} &  \theta_{52} &  \theta_{53} &  \theta_{54} &\theta_{55} &  \theta_{56} &  \theta_{57} &  \dots & \theta_{5N} \\
		\theta_{61} &  \theta_{62} &  \theta_{63} &  \theta_{64} &\theta_{65} &  \theta_{66} &  \theta_{67} &  \dots & \theta_{6N} \\
		\theta_{71} &  \theta_{72} &  \theta_{73} &  \theta_{74} &\theta_{75} &  \theta_{76} &  \theta_{77} &   \dots & \theta_{7N} \\
		\vdots & \vdots & \vdots & \vdots & \vdots & \vdots & \vdots &\ddots &\vdots  \\
		\theta_{N1} &  \theta_{N2} &  \theta_{N3} &  \theta_{N4} &\theta_{N5} &  \theta_{N6} &  \theta_{N7} & \dots & \theta_{NN} \\
	};
	\begin{scope}[on background layer]
		\node[fit=(m-1-1)(m-3-3), draw=black!100, thick, dashed, fill=lightgray!50, rounded corners] {};
		\node[fit=(m-1-4)(m-3-5), draw=black!100, fill=white!50, rounded corners] {};
		\node[fit=(m-1-6)(m-3-9), draw=black!100, fill=white!50, rounded corners] {};
		\node[fit=(m-4-1)(m-5-3), draw=black!100, fill=white!100, rounded corners] {};
		\node[fit=(m-4-4)(m-5-5), draw=black!100, thick, dashed,  fill=lightgray!50, rounded corners] {};
		\node[fit=(m-4-6)(m-5-9), draw=black!100, fill=white!50, rounded corners] {};
		\node[fit=(m-6-1)(m-9-3), draw=black!100, fill=white!50, rounded corners] {};
		\node[fit=(m-6-4)(m-9-5), draw=black!100,  fill=white!50, rounded corners] {};
		\node[fit=(m-6-6)(m-9-9), draw=black!100, thick, dashed,  fill=lightgray!50, rounded corners] {};
		\node[fit=(m-1-1)(m-1-1), draw=black!100, fill=gray!100, rounded corners] {};
		\node[fit=(m-2-2)(m-2-2), draw=black!100, fill=gray!100, rounded corners] {};
		\node[fit=(m-3-3)(m-3-3), draw=black!100, fill=gray!100, rounded corners] {};
		\node[fit=(m-4-4)(m-4-4), draw=black!100, fill=gray!100, rounded corners] {};
		\node[fit=(m-5-5)(m-5-5), draw=black!100, fill=gray!100, rounded corners] {};
		\node[fit=(m-6-6)(m-6-6), draw=black!100, fill=gray!100, rounded corners] {};
		\node[fit=(m-7-7)(m-7-7), draw=black!100, fill=gray!100, rounded corners] {};
		\node[fit=(m-8-8)(m-8-8), draw=black!100, fill=gray!100, rounded corners] {};
		\node[fit=(m-9-9)(m-9-9), draw=black!100, fill=gray!100, rounded corners] {};
		
	\end{scope} ;  
\end{tikzpicture}
\normalsize \label{SectoralVFED} 
\end{equation}

\vspace{3mm}

\noindent where the dark gray, light gray, and white boxes denote, respectively,  own variance shares,  co-movement shares, and contagion shares. The own variance shares capture the fraction of the forecast-error variance of asset \(i\) due to shocks from asset $i$ itself; the co-movement shares capture
the fraction due to shocks from other assets in the same cluster; and the contagion shares capture the fraction due to shocks from other assets in other clusters. In contrast, computing VDs with orthogonalized impulse responses yields only own variance and contagion shares, and computing VDs with generalized impulse responses yields only own variance and co-movement shares. 

The distinction between co-movement and contagion stems from the correlation of residuals. Within a cluster, the structural residuals are correlated, so we are unable to pinpoint the shock to a given asset; instead we observe only  co-movement (light gray areas). However, due to the constraints imposed via \(\boldsymbol{Q}_C\), the structural residuals are uncorrelated across clusters. This means that we are able to narrow down the origin of a shock to a given cluster, and quantify the reverberation across clusters. The absence of cross-cluster correlation of the structural residuals and the resulting causality of spillovers translate into a quantifiable contagion across clusters (white areas).

\subsubsection{Connectedness Measurement Within and Across Clusters}
 
Following \cite{diebold2012better}, we now define empirical average connectedness measures that parallel the theoretical concepts sketched above. We define the  {own variance share} for a cluster \(c\), which captures how much of the forecast error variance of cluster \(c\) is due to shocks specific to that same cluster, as
$
\Theta_{c}^{own} = \frac{1}{N_{c}} \sum_{i \in c}  \theta_{ii},  \nonumber
$
where $N_c$ is the number of members of cluster $c$.

In addition to own variance shares, we also measure connectedness across assets, but unlike \cite{diebold2012better}, who use generalized IRFs to obtain VDs, our clustering framework allows us to distinguish between two types of connectedness: Co-movement (within-cluster) and contagion (cross-cluster).  {Co-movement shares} capture the extent to which the forecast-error variance of cluster \(c\) is driven by co-movements among assets in that cluster, $\Theta_{c}^{comove} = \frac{1}{N_{c}} \sum_{\substack{i,j \in c\\ i\neq j}} \theta_{ij}. \nonumber$ Alternatively, {contagion shares} capture the extent to which the forecast-error variance of cluster \(c\) is driven by shocks from another cluster \(k\), $\Theta_{c\leftarrow k}^{contag} = \frac{1}{N_{c}}\sum_{i \in c} \left(\frac{1}{N_{k}} \sum_{j \in k} \theta_{ij} \right). \nonumber$  The {total contagion} received by  cluster \(c\) is then $\Theta_{c\leftarrow \bullet}^{contag} = \sum_{k\neq c} \Theta_{c\leftarrow k}^{contag}. \nonumber$  It will also prove useful to consider cross-cluster averages of the above own, co-movement, and total contagion measures. We write $\Theta^{own} = \frac{1}{C}\sum_{c=1}^C  \Theta_{c}^{own}$, $\Theta^{comove} = \frac{1}{C}\sum_{c=1}^C  \Theta_{c}^{comove}$, and $\Theta^{contag} = \frac{1}{C}\sum_{c=1}^C \Theta_{c\leftarrow \bullet}^{contag}.$  Note that \(\Theta^{own} + \Theta^{comove} + \Theta^{contag} = 1\).

\section{Clustered Connectedness in Global Equity Markets} \label{stocks}

Global equity markets are likely connected both locally (within regions) and globally (across regions), but the strengths and directions of connectedness are generally unknown, and moreover, they may be time-varying.  Simultaneously, improved quantitative characterization of market network connectedness would be of value not only to academic economists (of course), but also to a variety of financial-market participants, including private-sector agents (e.g., for improved portfolio allocation, risk management, and business planning), policymakers (e.g., for improved anticipation and tracking of cross-market spillover episodes as, for example, in the financial crises of 2007-9), and regulators (e.g., for improved monitoring of the effects of balance sheet and other linkages among financial institutions and trading exchanges). Additional beneficiaries include those not directly involved in financial markets, but who may nevertheless want to use the markets to help assess the effects of non-financial policies, such as imposition of tariffs or sanctions).

 Before proceeding, however, we emphasize that our approach is intentionally non-structural. Rather, it is meant to be a flexible reduced-form description of a set of covariance-stationary series, subject only to the assumed block-diagonality of the underlying structural shock covariance matrix, \textit{consistent with whatever (generally unknown and potentially time-varying) structural DGP is operative}. Relative to a more structural approach, our approach has both potential benefits (no damage from imposing false asset-pricing theory restrictions) and potential costs (less efficiency from failing to impose true  asset-pricing theory restrictions). We feel that the benefits of our approach significantly outweigh the costs, particularly given the strikingly poor empirical performance of structural asset-pricing models in equity premium prediction, as in the classic work of \cite{GoyalandWelch2008}.
 
Factor structure, for example, is one such popular asset-pricing theory restriction (assumption), which again may or may not be valid. It implies restrictions on a reduced-form VAR and is therefore compatible with, and encompassed by, a VAR approach, as shown by \cite{SW2005}. If linear factor structure holds, then a VAR will capture it, and if one \textit{knows} that such structure holds, then imposing it on the VAR may produce some efficiency gains, but one never knows. And even within the narrow world of linear factor models there are many possibilities -- single-factor, multi-factor, strong, weak, exact, approximate, etc.  
     
 Against this background, in this section we use our clustering framework to study connectedness in sixteen country equity markets spanning three global regions.  We proceed as follows. In section  \ref{full} we discuss estimation of the network-approximating VAR and provide full-sample analyses. In section \ref{rolling} we provide rolling-sample analyses. Finally, in section \ref{disc2} we discuss differences in results under clustered vs generalized identification. 
 
\subsection{Full-Sample Connectedness} \label{full}

Here we characterize country equity market connectedness using the full data sample.  If the structure of connectedness is fixed over time, then full-sample estimation is of immediate and unique interest, and even if it varies over time, the full-sample estimates provide a ``time-averaged" or ``unconditional" summary. Later, in section \ref{rolling}, we will explicitly allow for  time-variation in conditional connectedness via rolling-sample analysis. 

\subsubsection{Country Return Data and Network VAR Estimation} \label{data}

Our sample includes sixteen countries spanning three global regions: North America (U.S. and Canada), Europe (Germany, France, United Kingdom, Portugal, Italy, Ireland, Greece, and Spain), and East Asia (Japan, China, South Korea, Taiwan, Hong Kong, and Singapore). The sample period is 10 July 2002 (corresponding roughly to the full launch of the Euro) through 29 December 2021.

We construct the sixteen series of weekly nominal local-currency equity market returns. Our use of local returns is intentional and reflects common practice among portfolio managers, who often invest internationally while intentionally failing to fully hedge currency exposure. The reason is that although currency risk is potentially an important factor in long-horizon risk-adjusted returns, the hedging decision also  depends on a variety of other factors, including hedging costs, risk tolerance, investment objectives, and investment horizon. In addition, as a pragmatic matter, although currencies obviously fluctuate significantly over long horizons, their fluctuations have little impact on higher-frequency (daily or weekly) stock returns.

We proceed by taking daily  local-currency equity market indices, $P_t$, from the Wharton Research Data Services (WRDS) database, and   converting them to daily (log) returns $r_t$,  using $r_t = \Delta \log P_t$.\footnote{See \url{https://wrds-www.wharton.upenn.edu/}.} Next, we convert the returns from daily to weekly by cumulating the daily returns from Thursday through Wednesday each week. We use Thursday to Wednesday to avoid distortions due to beginning-of-week and end-of-week trades.  

In Table \ref{sumstats} we provide return summary statistics, grouped by region, which in this application we naturally take to be the relevant continent. Mean returns across markets  are sometimes positive and sometimes negative, but generally near zero, with standard deviations much larger.  Skewnesses, like means, are small and of mixed sign, whereas kurtoses are generally larger and well above three, consistent with the well-known fat tails in high- and medium-frequency asset returns.

\begin{table}[tbp]
	\caption{Summary Statistics\\Sixteen Weekly Country Equity Market Returns} \label{sumstats}
	\vspace{-3mm}
	\begin{center}
		\begin{tabular}{lllrrrrr} 
			\toprule 
			Region & Country & Label & Mean & Std & Info & Skew & Kurt\\  
			\midrule 
			North America &  All & & 6.96 & 15.60 & 0.45 & -1.42 & 12.74 \\ 
			North America &  United States &  USA &  8.25 & 16.65 & 0.50 & -1.02 & 10.18 \\ 
			North America &  Canada &  CAN &  5.68 & 16.09 & 0.35 & -1.52 & 15.14 \\ 
			\midrule 
			Europe &  All & & 1.17 & 18.44 & 0.06 & -0.98 & 9.07 \\ 
			Europe &  France &  FRA &  4.91 & 19.75 & 0.25 & -0.69 & 10.24 \\ 
			Europe &  Germany &  GER &  4.43 & 19.41 & 0.23 & -1.00 & 9.97 \\ 
			Europe &  United Kingdom &  GBR &  2.85 & 16.47 & 0.17 & -0.79 & 9.40 \\ 
			Europe &  Portugal &  PRT &  -0.36 & 19.60 & -0.02 & -1.03 & 8.66 \\ 
			Europe &  Ireland &  IRL &  3.76 & 23.79 & 0.16 & -0.95 & 10.86 \\ 
			Europe &  Italy &  ITA &  0.39 & 21.18 & 0.02 & -0.76 & 7.71 \\ 
			Europe &  Greece &  GRC &  -8.32 & 31.06 & -0.27 & -0.66 & 8.89 \\ 
			Europe &  Spain &  SPA &  1.69 & 20.56 & 0.08 & -0.47 & 6.70 \\ 
			\midrule 
			East Asia &  All & & 4.79 & 15.52 & 0.31 & -0.83 & 8.97 \\ 
			East Asia &  Japan &  JPN &  3.74 & 19.78 & 0.19 & -0.70 & 8.13 \\ 
			East Asia &  China &  CHN &  5.04 & 24.59 & 0.20 & -0.69 & 8.08 \\ 
			East Asia &  South Korea &  KOR &  6.71 & 20.48 & 0.33 & -0.60 & 10.02 \\ 
			East Asia &  Taiwan &  TWA &  5.61 & 19.05 & 0.29 & -0.58 & 7.49 \\ 
			East Asia &  Hong Kong &  HKG &  4.97 & 20.52 & 0.24 & -0.46 & 7.12 \\ 
			East Asia &  Singapore &  SGP &  2.66 & 15.85 & 0.17 & -0.76 & 10.87 \\ 
			\midrule 
			Global &  All & & 3.25 & 15.60 & 0.21 & -1.16 & 10.06 \\ 
			\bottomrule 
		\end{tabular}
	\end{center}
	\footnotesize
%	\vspace{-3mm}   
\begin{spacing}{1}
Notes: We present summary statistics for annualized weekly local-currency nominal equity returns for sixteen country markets. The weekly returns are from Thursday to Wednesday, and the sample period is 10 July 2002 through 29 December 2021.  ``Mean" denotes the (sample) mean, ``Std" denotes standard deviation,  ``Info" denotes the information ratio (Mean/Std), ``Skew" denotes skewness, and ``Kurt" denotes kurtosis.  See text for details.
\end{spacing}		
\normalsize
\end{table}

Following \cite{demirer2018estimating}, we proceed with equation-by-equation estimation of a 16-variable VAR(3) using an adaptive elastic net \citep{zou2009adaptive} for regularization (shrinkage and selection), which is generally preferred to related penalized estimators (e.g., lasso \citep{tibshirani1996regression}) in environments like ours with correlated predictors and grouped structure.  In particular, for each equation we solve 
\begin{equation}
	\boldsymbol{\widehat{\beta}} = {\rm argmin}_{\boldsymbol{\beta}} \left ( \sum_{t=1}^T \left (y_t - \sum_i \beta_i x_{it} \right )^2 + \lambda \sum_{i=1}^K w_i \left (   \frac{1}{2}    |{\beta_i}| + \frac{1}{2} \beta_i^2 \right ) \right ), \label{AENeqn}
\end{equation}
where $w_i = 1 / {|\hat{\beta}_{i,OLS}|}$ and $\lambda$ is selected equation-by-equation by 10-fold cross-validation.\footnote{The adaptive elastic net penalty averages the LASSO penalty with a ridge penalty, and moreover it weights the average by inverse OLS parameter estimates, thereby  shrinking the smallest OLS-estimated coefficients most heavily toward zero.} Once all equations of the VAR have been estimated, we  obtain  the residuals for each equation, from which we  obtain in the usual way an estimate $\widehat{\boldsymbol{\Sigma}}$ of the reduced-form shock covariance matrix.

We assume that clustering is by region.  Market ordering within the three clusters is irrelevant for estimated VD network structure, but cluster ordering is potentially relevant, and there are \(3! = 6\) possible cluster orderings. For a given cluster ordering,  the appropriate ``clusterizing" (as opposed to ``orthogonalizing") transformation matrix $\widehat{\boldsymbol{Q}}^{-1}_3$ is 
\begin{equation}
	\widehat{{\boldsymbol{Q}}}^{-1}_3 = 
	\begin{pmatrix}
		\boldsymbol{I} &0 &0\\
		-\widehat{\boldsymbol{\Sigma}}_{21} \widehat{\boldsymbol{\Sigma}}_{11}^{-1}& \boldsymbol{I}  &0 \\
		- \widehat{\boldsymbol{\Sigma}}_{31} \widehat{\boldsymbol{\Sigma}}^{11}- \widehat{\boldsymbol{\Sigma}}_{32} \widehat{\boldsymbol{\Sigma}}^{21}&-\widehat{\boldsymbol{\Sigma}}_{31} \widehat{\boldsymbol{\Sigma}}^{12}-\widehat{\boldsymbol{\Sigma}}_{32}\widehat{\boldsymbol{\Sigma}}^{22}& \boldsymbol{I}
	\end{pmatrix},
\end{equation}
as per equation \eqref{clusterize} and Appendix \ref{Q3}. We then obtain VDs from the clusterized VAR, using a VD horizon of $h=12$ weeks. There are of course six cluster orderings that can be used to produce clusterized VDs. In an effort to remain agnostic regarding orderings, we calculate VDs for all possible orderings and report averages. 

\subsubsection{The Market Network Graph with Clustered Identification} 
  
 \begin{figure}[tbp]  	
 	\caption{Estimated Global Equity Market Return Network Graph, Clustered Identification}  	\label{fig:fsnetwork_clustered}
 	\begin{center}
%\begin{adjustwidth}{-3mm}{0mm}
 	\includegraphics[scale=.3, trim={0cm 0cm 0cm 5mm},clip]{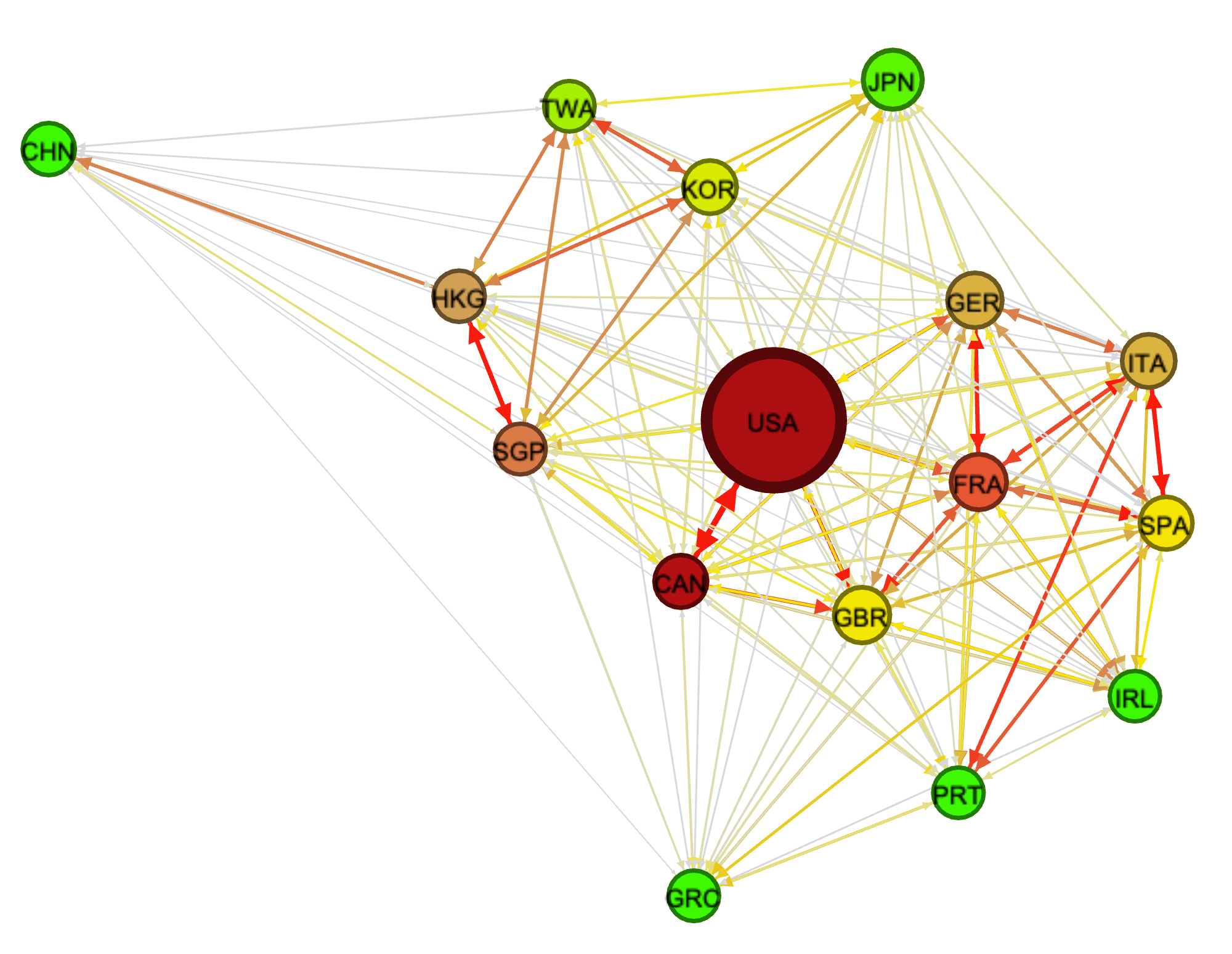}
%\end{adjustwidth}
 	 \end{center}
\footnotesize
\vspace{-2mm}  
\begin{spacing}{1}
Notes: We show the estimated network spring graph obtained from the ForceAtlas2 algorithm.  See text for details.
\end{spacing}
\normalsize
 \end{figure}

%	\vspace{10mm}

\begin{figure}[tbp]
	\caption{Cross-Country Total Directional Connectedness Densities, Clustered Identification} \label{dens1}
	\begin{center}
		\includegraphics[scale=.8, trim={0cm 0mm 0cm 17mm},clip]{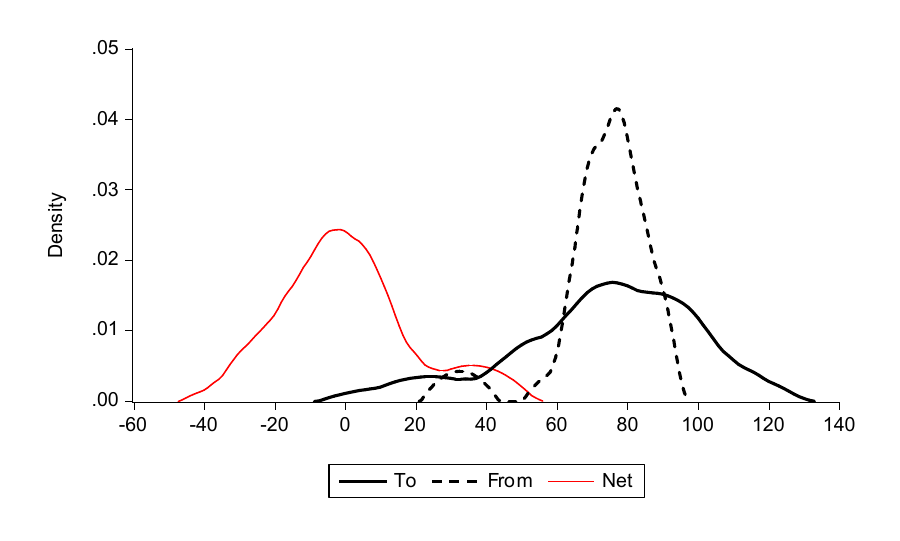}
	\end{center}
	\footnotesize
	\vspace{-7mm}  
	\begin{spacing}{1}
	Notes: We show full-sample kernel density estimates for three total directional connectedness measures (to, from, and net) across sixteen country equity markets, using clustered identification. The ``to" density is solid black, the ``from" density is dashed black, and the ``net" density is red.  See text for details.
	\end{spacing}
	\normalsize
\end{figure}

\begin{figure}[tbp]  	
	\caption{Estimated Global Equity Market Return Network Graph, Generalized Identification} 	\label{fig:fsnetwork_generalized}
	%\begin{adjustwidth}{-3mm}{0mm}
	\begin{center}
	\includegraphics[scale=.25, trim={0cm 0cm 0cm 5mm},clip]{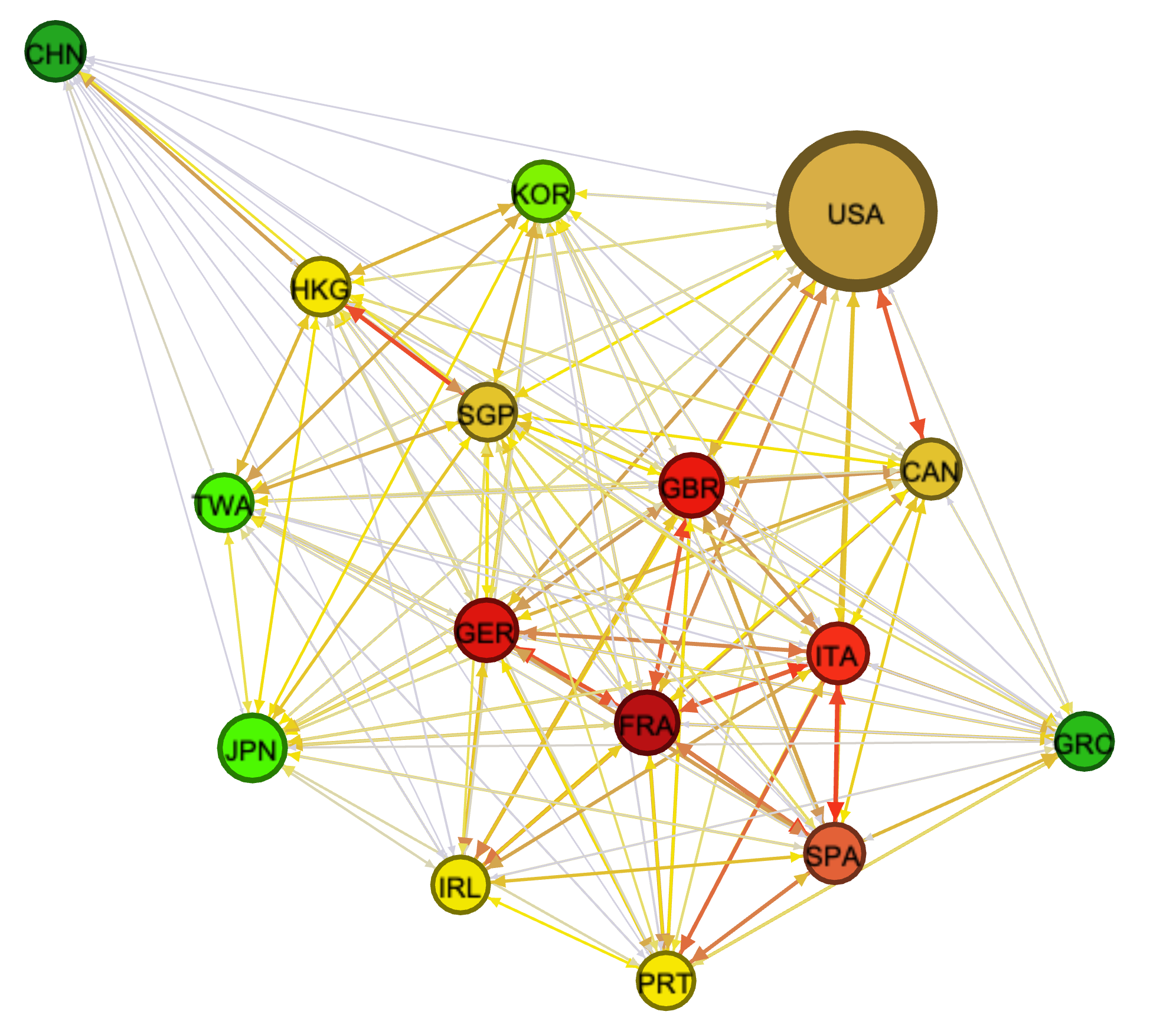}
	\end{center}
	%\end{adjustwidth}
	\footnotesize
	\vspace{-2mm}  
	\begin{spacing}{1}
		Notes: We show the estimated network spring graph obtained from the ForceAtlas2 algorithm.  See text for details.
	\end{spacing}
	\normalsize
\end{figure}

\begin{figure}[tbp]
	\caption{Cross-Country Total Directional Connectedness Densities, Generalized Identification} \label{dens2}
	\begin{center}
		\includegraphics[scale=.8, trim={0cm 0mm 0cm 7mm},clip]{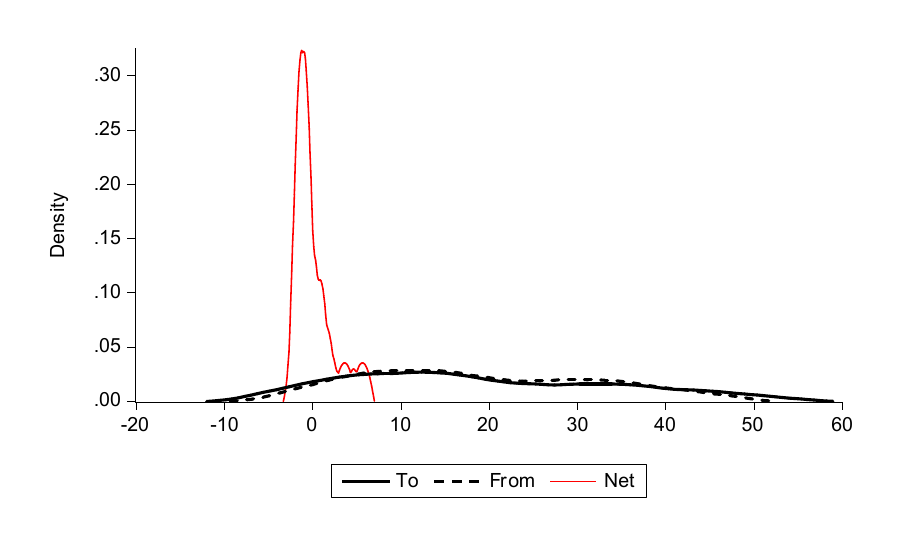}
	\end{center}
	\footnotesize
	\vspace{-7mm}  
	\begin{spacing}{1}
		Notes: We show full-sample kernel density estimates for three total directional connectedness measures (to, from, and net) across sixteen country equity markets, using generalized identification.   The ``to" density is solid black, the ``from" density is dashed black, and the ``net" density is red.  See text for details.
	\end{spacing}
	\normalsize
\end{figure}

We visualize estimated networks using ``spring graphs" obtained from the ForceAtlas2 algorithm of \cite{jacomy2014forceatlas2}, as implemented in the open-source Gephi software.\footnote{See  \url{https://gephi.github.io/}.} The algorithm finds a steady state in which repelling and attracting forces exactly balance, where nodes repel each other like similar poles of two magnets, while edges (links), attract their nodes like springs, with the attracting force proportional to average pairwise directional connectedness ``to" and ``from."\footnote{Steady state node locations depend on initial node locations and are therefore not unique, but that is largely irrelevant for us, as we are interested in relative, not absolute, node locations in equilibrium.}  
  
 There are six associated graph components: Node label, node size, node color, edge thickness, edge color, and edge arrow size (two per edge, because the network is directed).\footnote{For details see \cite*{demirer2018estimating}.} Node label indicates the country as shown in Table \ref{sumstats}. Node size indicates the total capitalization of the country's equity market, obtained from the World Bank's World Development Indicators database.  Node color, very importantly, indicates the net directional connectedness to others, ranging from bright green, the weakest, to vivid yellow, to brick yellow, to bright red, to dark red, the strongest.  Edge thickness indicates the average directional pairwise connectedness between two nodes.  As it is not always easy to discern the thickness difference between two edges, we also use edge color to indicate the average directional pairwise connectedness between two nodes. Edge color follows a similar scale to node color, starting with light gray, the weakest, followed by vivid yellow, brick yellow, and red, the strongest. Edge arrow size, also very importantly, indicates the pairwise directional connectedness from one node to the other.

In Figure \ref{fig:fsnetwork_clustered} we present the network spring graph for our clustered identification.  Several aspects of the network are apparent. First,  North America clusters together, Europe clusters together, and East Asia clusters together, but with two major outliers: GRC for Europe and CHN for East Asia. Alternatively, another three-cluster interpretation could be: Anglo-American (USA, CAN, GBR), ``Core Europe" (GER, FRA, SPA, ITA, IRL, PRT), and ``Core East Asia" (SGP, HKG, KOR, TWA, JPN), again with GRC and CHN as outliers.
	
Second, regardless of which interpretation one adopts, GBR plays a key role in linking North America and Europe.  In particular, both USA and CAN have strong pairwise directional connectedness to GBR, which then links strongly to FRA and onward to the rest of Europe (except GRC).  (There are also strong directional links (large arrows) from USA to GER, FRA, and ITA.)
		
Finally,  most obviously and importantly,  North America sits squarely in the network graph center, with red nodes indicating very high net directional connectedness to others. That is, on balance  North America sends large amounts of 12-week-ahead uncertainty to others. We have mentioned already the strong pairwise directional connectedness from  North America to the European countries of GBR, FRA, GER, and ITA, and there is similarly strong directional connectedness (large, if not red, arrows) from  North America to all East Asian countries except CHN. 	
	
In closing this section, we highlight an additional important aspect of the country equity market network graph: Total directional connectedness ``to" others and ``from" others (and their difference,  ``net" total directional connectedness). The total directional measures are of course implicit in the spring graph of Figure  \ref{fig:fsnetwork_clustered}, which provides a complete network characterization, but it is impossible to extract them visually.  Hence in Figure \ref{dens1} we supplement the spring graph with estimates of the cross-country densities of total directional connectedness (to, from, and net).  The ``to" density has a similar mean but is more dispersed than the ``from" density; that is, ``uncertainty transmissions" range more widely across countries than do ``uncertainty receipts". In addition, both the ``to" and ``from" densities are skewed left; indeed the ``from" density has a small second left mode.  Hence the ``net" (``to" minus ``from"; that is, net transmissions) density is centered near zero but skewed right, with a second right mode corresponding to a few countries with large net transmissions (USA, Canada, and France).

\subsubsection{Benchmarking Clustered Identification}

For comparison to Figure \ref{fig:fsnetwork_clustered}, we show the network graph for generalized as opposed to clustered identification in Figure \ref{fig:fsnetwork_generalized}.  The two graphs have both similarities and differences.  Let us begin with similarities. In both graphs there are clear connectedness clusters for Europe and East Asia, with countries in both regions located closely together.  In both graphs GRC and CHN are outliers, indicating relatively weak connections to other equity markets, including those in their own regions of Europe and East Asia, respectively.  Finally, in both graphs GRC and CHN feature bright green nodes, indicating that they are net recipients of future uncertainty from other countries.

Now let us consider differences between the generalized and clustered network graphs.  Most importantly and obviously,  North  America is located on the outskirts of the generalized identification graph, which indicates that it has only weak links to other equity markets and, hence, a marginal role in the global equity market return  network. Furthermore, its brown node also indicates that its net connectedness to other markets is rather low.  Recall that, in contrast, in the clustered identification graph  North America sits at the center, with a red node, indicating that it plays a crucial role in the global return network -- indeed it is the most significant generator of net return connectedness among all markets.

Finally, we show total directional connectedness densities under generalized identification in Figure \ref{dens2}, for comparison to the densities under clustered identification in Figure \ref{dens1}. The situation under generalized identification differs greatly. In particular, the ``to" density (``uncertainty transmissions") and ``from" density (``uncertainty receipts") are very similar under generalized identification, which makes the ``net" (transmissions) density tightly centered around zero. That is, under generalized identification no countries are identified as disproportionately large net transmitters, in contrast to the clear identification of USA, CAN, and FRA as large net transmitters under clustered identification.

\subsection{Rolling-Sample Connectedness} \label{rolling}

We now allow for time-variation in connectedness by changing from full-sample estimation to rolling-sample estimation. The window width for rolling-sample estimation is 2 years (104 weeks).  In addition, it will now prove useful to present results for various aspects of clustered connectedness (CC) and generalized connectedness (GC) simultaneously, rather than sequentially as we did for full-sample estimation.

\subsubsection{System-Wide Connectedness and its Components}

\begin{figure}[tbp]
	\caption{System-Wide Connectedness and its Components}  	\label{fig:gcind_within&across}
	\begin{center}
		(a) Clustered Identification\\
		\includegraphics[scale=.650,trim = 0cm 0cm 0cm 8mm, clip]{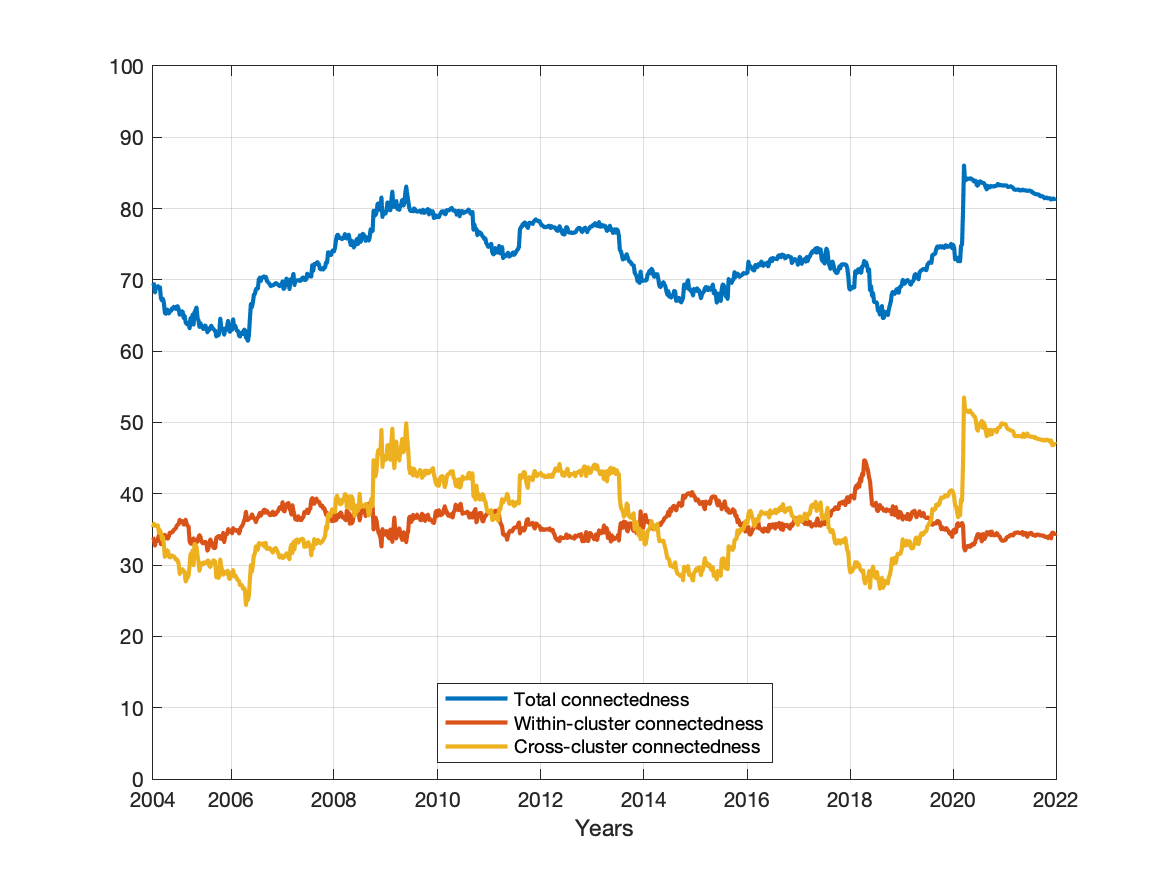}\\
		\vspace{5mm}
		(b)  Generalized Identification\\
		\includegraphics[scale=.650,trim = 0cm 0cm 0mm 8mm, clip]{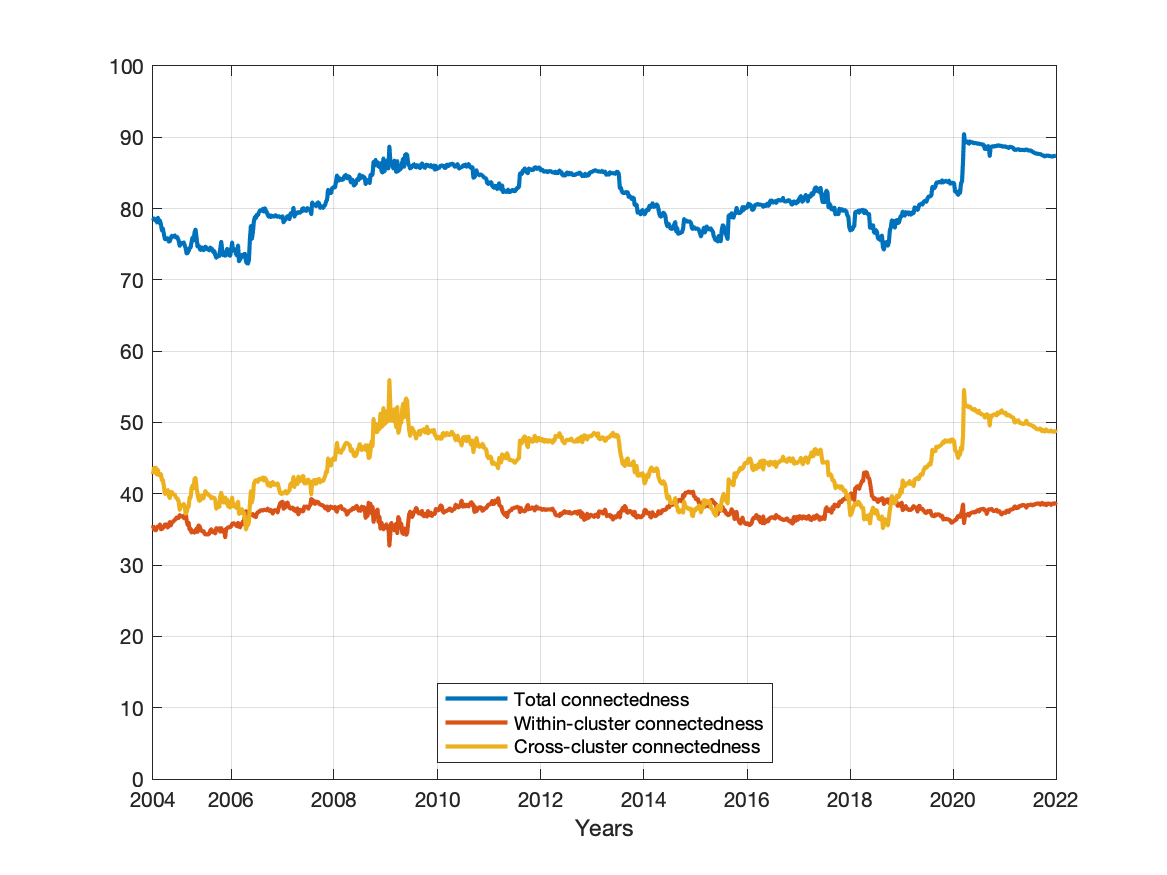}\\ 
	\end{center}
\end{figure}

In Figure \ref{fig:gcind_within&across} we show system-wide connectedness (the sum of all off-diagonal VD matrix elements), together with its within-cluster and cross-cluster components (sums of all off-diagonal VD matrix elements inside clusters and outside clusters, respectively). 

Let us first discuss the CC results in the upper panel of Figure \ref{fig:gcind_within&across}. System-wide CC has two prominent movements, first a large and multi-year increase in 2007-2009 during the global financial crisis, and then a large sharp increase in early 2020 as the global pandemic emerged.  Examination of the within- and cross-cluster components reveals that the system-wide CC movements are driven largely by clear and pronounced movements in the cross-cluster component.  The within-cluster component, in contrast, is quite stable.

Now let us compare the just-discussed upper-panel CC results to the lower-panel GC results. The system-wide CC and GC movements are clearly very similar, with CC always below GC. This is expected, because the GC approach allows for simultaneous shocks to all variables, both within and across regions, whereas the CC approach imposes uncorrelated shocks across regions, as we discuss in greater detail in section \ref{disc2} below. System-wide CC and GC are closest during the global financial crisis (following the collapse of Lehman Brothers in mid-September 2008) and the spread of the COVID-19 pandemic to the West (in early March 2020). Closer inspection, however, reveals a key difference between the CC and GC measures: Movements in CC are sharper and more pronounced than those of GC, with system-wide CC (and its key driver, cross-cluster CC) varying over a wider range.

\subsubsection{Regional Net Directional Connectedness}

\begin{figure}[tbp]
	\caption{Regional Net Directional  Connectedness}  \label{fig:regnetconn}
	\begin{center}
		(a) Clustered Identification\\
		\includegraphics[scale=0.65,trim = 0cm 0cm 0cm 8mm, clip]{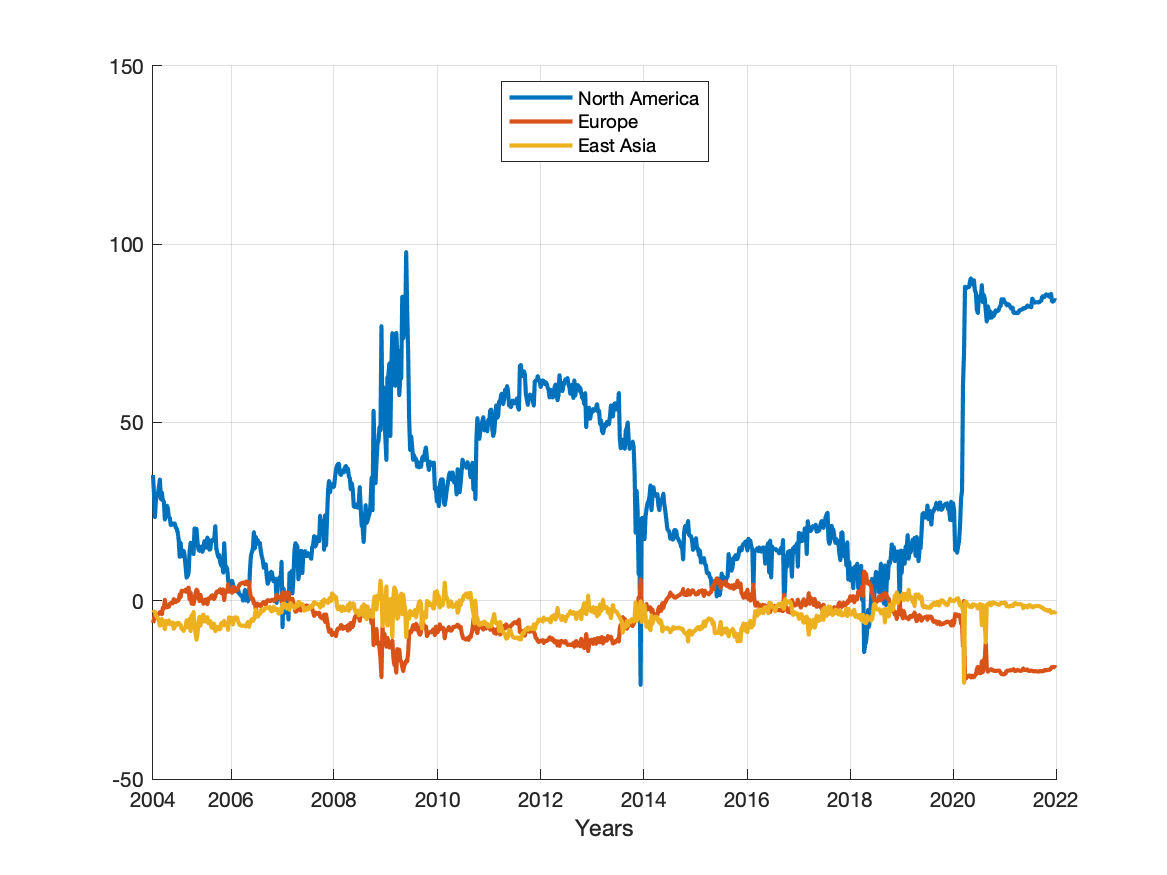}\\
		\vspace{5mm}
		(b) Generalized Identification\\
		\includegraphics[scale=0.65,trim = 0cm 0cm 0cm 8mm, clip]{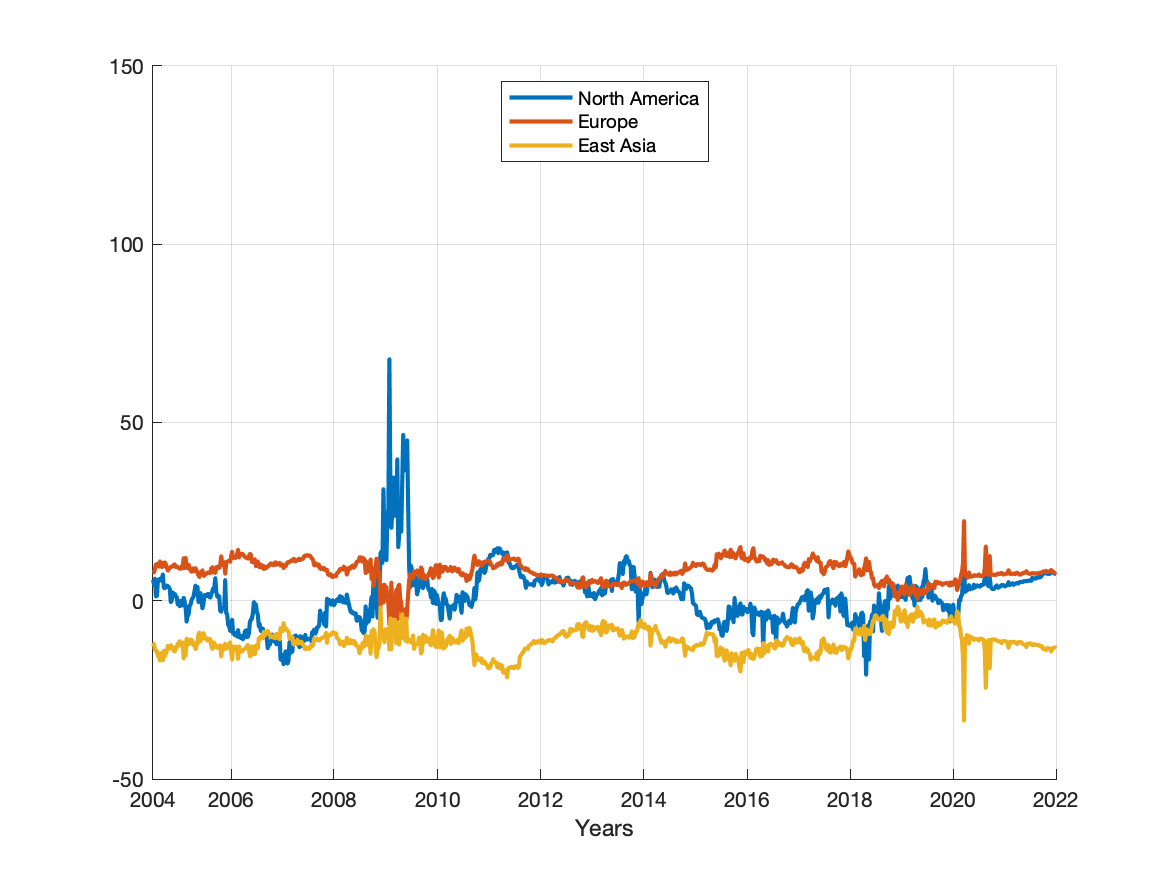}\\
	\end{center}
\end{figure}

In Figure \ref{fig:regnetconn} we show regional net  directional connectedness ``to" others (that is, net transmissions of future uncertainty from one region to the other two -- the sum of all out-of-region VD matrix elements in the region's columns, minus the sum of all out-of-region VD matrix elements in the region's rows), for the North America, Europe, and East Asia regions.\footnote{We normalize by the number of countries in the transmitting region.} The top panel of the figure is based on cluster identification (regional net  CC), and the bottom panel is based on generalized identification (regional net GC).

Let us first consider the regional net CC shown in the top panel of Figure \ref{fig:regnetconn}, starting with North America. Throughout the sample, North American net transmissions to Europe and East Asia are positive and typically very large (and often huge) relative to European and East Asian net transmissions to North America. Key episodes include: 

\vspace{-2mm}

\begin{enumerate}

	\item The financial crisis of 2007-2009. Following the summer 2007 escalation of tensions in the U.S. mortgage and financial markets, North American net CC climbed significantly by the late 2007 and surged following the collapse of Lehman Brothers in late 2008, peaking in early 2009 before dropping.

	\item The financial crises of 2010-2014.  As the North American crisis moved to Europe, it created a hump-shaped  North American net transmissions trajectory that started in 2010, peaked in 2011-2012, and subsided by 2015, linked to the series of European crises in Greece, Portugal, Ireland, Iceland, Italy, and Spain, with the North American transmissions absorbed almost exclusively by Europe.

	\item The onset of the COVID-19 pandemic in 2020.  COVID-19 burst into the Western Hemisphere in early 2020, producing a huge increase in North American net transmissions.
 
\end{enumerate}

\vspace{-2mm}

\noindent Europe and East Asia, to which we now turn, had very different net CC experiences. 

In contrast to the typically large, positive, and fluctuating values of North American net CC, European and East Asian net CC are typically small, negative, and stable. Europe and East Asia are largely net recipients of transmissions from North America (i.e., they have negative net CC). European net receipts, for example, increase sharply (i.e., European net CC decreases sharply, becoming even more negative) during the major North American net transmissions episodes sketched above.
 
Now let us compare the just-discussed regional net CC results to the GC results in the lower panel of Figure \ref{fig:regnetconn}.  In general the movements in GC are less pronounced than those of CC, particularly for North America, just as was the case earlier for system-wide connectedness and its components in Figure \ref{fig:gcind_within&across}. The 2020 pandemic outbreak, for example, is hardly noticeable in GC North American net transmissions.

Moreover, there are important GC vs CC differences in regional net transmissions well beyond the lower resolution of North American GC movements. In particular, North American transmissions fluctuate around zero under the GC, Europe's are consistently positive, and East Asia's are consistently negative.

\subsection{On Connectedness Under Clustered vs Generalized Identification} \label{disc2}
 
\begin{figure}[tbp]  	
	\caption{Correlation Matrix,\\Sixteen Weekly Country Equity Market Returns} 	\label{fig:pairwise correlation}
	\vspace{-7mm}
	\begin{center}
		\includegraphics[scale=.4, trim={0cm 0cm 0cm 15mm},clip]{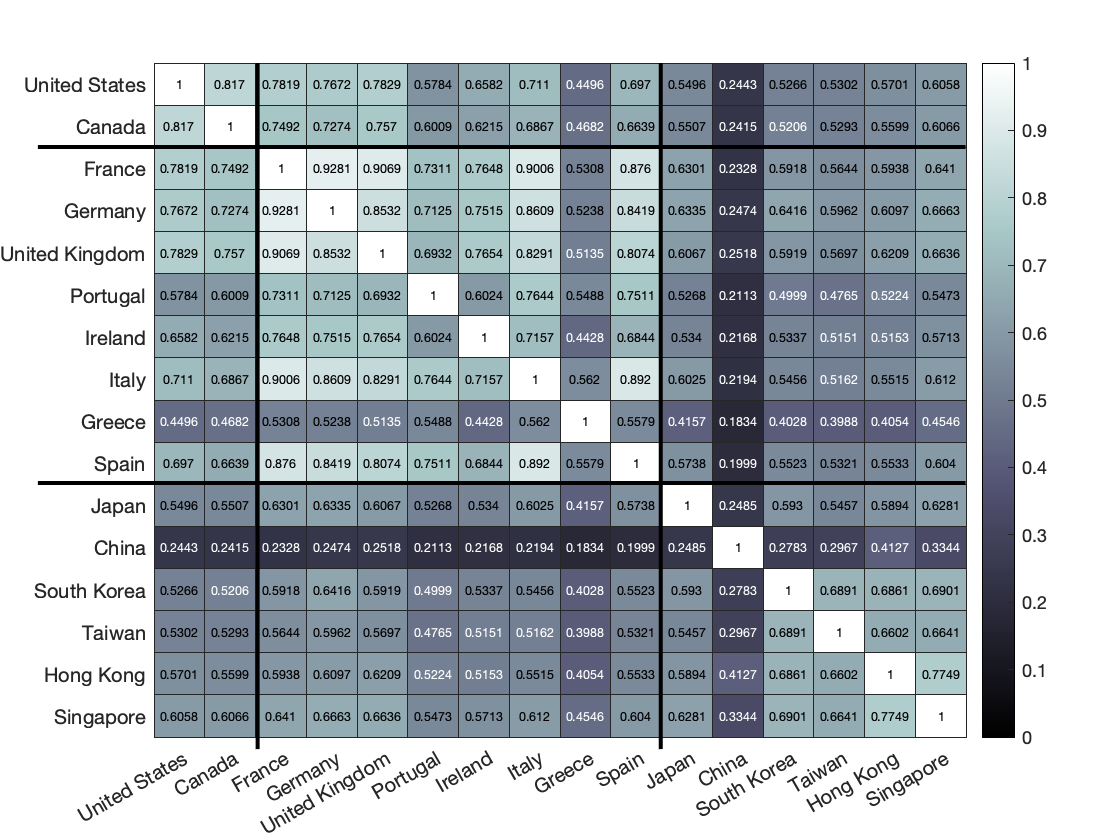}
	\end{center}
	\footnotesize 
	\vspace{-2mm}  
	\begin{spacing}{1}
	Notes to figure: Shading indicates strength of correlation, with brighter  shades indicating higher correlation.
	\end{spacing}
	\normalsize
\end{figure}

\begin{figure}[tb]
	\caption{Differences in Connectedness\\Generalized Minus Clustered Identification}  \label{fig:delta_con}
	\begin{center}
		\vspace{-5mm}
		\includegraphics[scale=0.7, trim={0cm 0cm 0cm 10mm},clip]{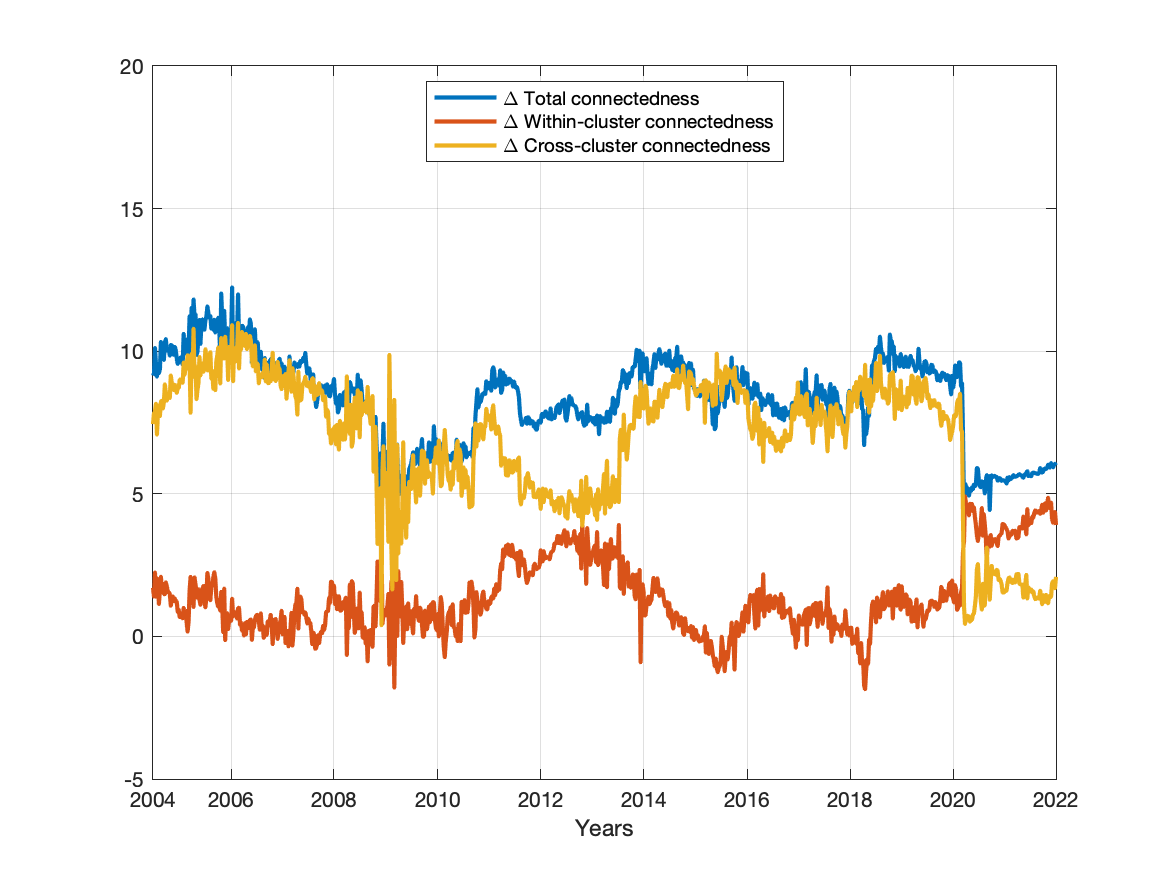}
	\end{center}
	\footnotesize 
	\vspace{-5mm}  
	\begin{spacing}{1}
		Notes to figure: We show the changes in connectedness (total system-wide, within-cluster, and cross-cluster) when moving from clustered identification to generalized identification. 
	\end{spacing}
	\normalsize
	
\end{figure}

Having provided results for both full-sample and rolling-sample clustered connectedness, and having emphasized differences under clustered vs generalized identification, we now provide some additional insight into the reasons for the differences, the essence of which is that generalized identification captures co-movement across nodes, but not contagion. As such, both within- and cross-cluster connectedness under generalized identification summarize aspects of correlation, but not causality, so that the clustered and generalized approaches can produce very different results. 
 
 In the absence of orthogonalization across regions (i.e., under generalized identification), shocks are subject to a feedback loop that smooths them across the system, so that shocks cannot be properly attributed to their origin.\footnote{Closely related, in the absence of orthogonalization across regions, the densities of the ``from" and ``to" directional connectedness measures become diffuse and difficult to distinguish, as is clear from  comparing their shapes under clustered identification in Figure \ref{dens1} to those under generalized identification in Figure \ref{dens2}.} To see why, note that USA appears to be less central to the global equity markets than GER, FRA, SPA, and ITA in the generalized identification network graph of Figure \ref{fig:fsnetwork_generalized}, and simultaneously that those four European countries have the highest pairwise correlations in the dataset as shown in the correlation matrix of Figure \ref{fig:pairwise correlation}. What happens is that, in the absence of orthogonalization across regions, the high correlations in Europe capture the otherwise causal connectedness of USA.  In contrast, the orthogonalization across regions embedded in the clustered approach links co-movement to  within-cluster connectedness and contagion to cross-cluster connectedness, in which case USA emerges, as expected, as the key player in global equity markets as shown in Figure \ref{fig:fsnetwork_clustered}.

We show in Figure \ref{fig:delta_con} that total connectedness is always greater under generalized identification; that is, the total connectedness difference -- ``total generalized minus total clustered" -- is always positive. The reason is that shocks reverberate more across the system when not orthogonalized by region, translating into greater connectedness. Most of the total difference stems from the underlying cross-cluster difference, which again is always positive, because cross-cluster connectedness under generalized identification reflects co-movement in addition to contagion. In contrast, the within-cluster difference is typically near zero, with two key exceptions: the European debt crises of 2010-2014 and the emergence of COVID-19 in 2020. In each case, the within-cluster difference rises  but the cross-cluster difference falls.

\section{Summary and Directions for Future Research} \label{concl}

Network connectedness and its evolution are central in economics and finance, and a large literature has arisen that  explores connectedness measurement based on variance decompositions from VARs. However, those VARs are typically identified using full orthogonalization or no orthogonalization, which, although useful, are special and extreme cases of the more general and empirically-realistic ``clustered orthogonalization" approach developed in this paper, which allows for correlated structural shocks within clusters (e.g., asset classes, industries, regions, etc.) while maintaining orthogonality across clusters, thereby facilitating a nuanced empirical exploration of the ``contagion vs co-movement" distinction emphasized by \cite{forbes2002no}.

We used our clustered-connectedness framework to explore linkages in global equity returns for sixteen countries in three regions (North America, Europe, and East Asia). There are several key results. First, we identified major players (large net senders of future uncertainty) on both global and local scales.  Under clustered identification, the two North American countries, USA and CAN, are the largest global net senders by far, and FRA is a key European net sender. 

Second, we documented important time variation in connectedness.  Under  clustered identification, system-wide connectedness varies importantly, and its cross-cluster (as opposed to within-cluster) component is responsible for most of the variation. Regional net directional connectedness also varies importantly for the key regional net sender, North America, but much less so for the Europe and East Asia.

Finally, we found important differences in connectedness patterns for clustered vs generalized identifications, and we provided an explanation. Generalized identification is unable to uncover causal connections, instead attributing all node connections to simple correlation.

Now, in closing, let us mention some additional literature links and some promising directions for future research. Regarding additional literature links, it is interesting to note that the clustered network connectedness measurement methods introduced in this paper are part of a wave of recent econometric contributions addressing  measurement in other clustered contexts.  One prominent example is the now-large literature on estimation of panel data models with clustered fixed effects, beginning with \cite*{BonhommeManresa2015}.  Another prominent  example is estimation with clustered covariance matrices in both cross sections \citep*{MacKinnon2024} and time series  \citep*{Hansen2024}.

As for future research, one direction is empirical cluster classification, whether from a frequentist perspective in the tradition of \cite*{BonhommeManresa2022} and \cite*{ChiangSasakiWang2024}, or from a Bayesian perspective as in \cite{Zhang2024}. Of particular interest may be improved methods for detecting sender/receiver clusters, as in \cite*{Gudmundsson2021} and \cite*{Brownlees2022}. An elegant approach, incorporating not only clustering but also parameter variation, is provided by  \cite{LPTW2025}, who build on \cite{ZPLLW2017}.
	
An important related issue is cluster misclassification, regardless of whether the clusters are determined endogenously or specified exogenously. In this paper we have assumed that the DGP features clustering, and we have emphasized the network estimation distortions produced in that environment when non-clustered identification (most notably, generalized identification) is used. Conversely, however, clustered identification may produce distortions when clustering is imposed on a non-clustered DGP, or more generally when an adopted clustering pattern differs from the actual pattern.

%\clearpage
 
\appendix
\appendixpage
%\addappheadtotoc
 
\newcounter{saveeqn}
\setcounter{saveeqn}{\value{section}}
\renewcommand{\theequation}{\mbox{\Alph{saveeqn}.\arabic{equation}}} \setcounter{saveeqn}{1}
\setcounter{equation}{0}

\section{Cluster Orthogonalization by Linear Projection \label{Q3}}

Here we derive the matrix $\boldsymbol{Q}_C^{-1}$ such that transforming  the vector of VAR residuals $\boldsymbol{u}_t$ by  $\boldsymbol{Q}_C^{-1}$ orthogonalizes them across $C$  clusters. For clarity we display results for $C=3$; extension to $C>3$ is immediate but more tedious. That is, we seek $\boldsymbol{Q}_3^{-1}$ such that 
 \begin{equation}
	\mathbb{V}[\boldsymbol{Q}_3^{-1} \boldsymbol{u}_t]=
	\mathbb{V}[\boldsymbol{\epsilon}_t]=
	\boldsymbol{\Omega}=
	\begin{pmatrix}
		\boldsymbol{\Omega}_{11} & \textbf{0} &\textbf{0} \\
		\textbf{0}  & \boldsymbol{\Omega}_{22}  &\textbf{0} \\
		\textbf{0} & \textbf{0}  & \boldsymbol{\Omega}_{33}
	\end{pmatrix},  \nonumber
\end{equation}
where 
  \begin{equation}
	\mathbb{V}[\boldsymbol{u}_t]= \boldsymbol{\Sigma} = 
	\begin{pmatrix}
		\boldsymbol{\Sigma_{11}} & \boldsymbol{\Sigma_{12}}  &\boldsymbol{\Sigma_{13}}\\
		\boldsymbol{\Sigma_{21}} & \boldsymbol{\Sigma_{22}}  &\boldsymbol{\Sigma_{23}}\\
		\boldsymbol{\Sigma_{31}} & \boldsymbol{\Sigma_{32}} & \boldsymbol{\Sigma_{33}}
	\end{pmatrix},  \nonumber
\end{equation}
and $\boldsymbol{0}$ denotes a matrix of zeros.

We begin with $	\boldsymbol{\epsilon}_{1,t}$ and proceed sequentially, orthogonalizing residuals across clusters by linear projection, precisely as in the well-known Gram-Schmidt procedure:
  \begin{equation}
	\boldsymbol{\epsilon}_{1,t}
	=  \boldsymbol{u}_{1,t} \label{ortho1}
\end{equation}
\begin{equation}
	\boldsymbol{\epsilon}_{2,t}
	=  \boldsymbol{u}_{2,t}-\boldsymbol{\Sigma}_{21}\boldsymbol{\Sigma}_{11}^{-1}\boldsymbol{u}_{1,t} \label{ortho2}\\
\end{equation}
   \begin{align}
  \boldsymbol{\epsilon}_{3,t}
  &=  \boldsymbol{u}_{3,t}-
  \begin{pmatrix}
 \boldsymbol{\Sigma}_{31} &  \boldsymbol{\Sigma}_{32}
  \end{pmatrix}
    \begin{pmatrix}
  \boldsymbol{\Sigma}_{11} &  \boldsymbol{\Sigma}_{12} \\
    \boldsymbol{\Sigma}_{21} &  \boldsymbol{\Sigma}_{22}
  \end{pmatrix}^{-1}
   \begin{pmatrix}
   \boldsymbol{u}_{1t} \\    \boldsymbol{u}_{2,t}
  \end{pmatrix} \label{blockinverse} \\
  &=  \boldsymbol{u}_{3,t}-
  \begin{pmatrix}
  \boldsymbol{\Sigma}_{31} &  \boldsymbol{\Sigma}_{32}
  \end{pmatrix}
    \begin{pmatrix}
  \boldsymbol{\Sigma}^{11} &  \boldsymbol{\Sigma}^{12} \\
    \boldsymbol{\Sigma}^{21} &  \boldsymbol{\Sigma}^{22}
  \end{pmatrix}
   \begin{pmatrix}
   \boldsymbol{u}_{1,t} \\    \boldsymbol{u}_{2,t}
  \end{pmatrix} \nonumber \\
	 &=  \boldsymbol{u}_{3,t}-
  \begin{pmatrix}
 \boldsymbol{\Sigma}_{31}\boldsymbol{\Sigma}^{11}+\boldsymbol{\Sigma}_{32}\boldsymbol{\Sigma}^{21} &   \boldsymbol{\Sigma}_{31}\boldsymbol{\Sigma}^{12}+\boldsymbol{\Sigma}_{32}\boldsymbol{\Sigma}^{22}\\
  \end{pmatrix}  
   \begin{pmatrix}
   \boldsymbol{u}_{1,t} \\    \boldsymbol{u}_{2,t}  
  \end{pmatrix}, \nonumber
  \end{align}
where $\boldsymbol{\epsilon}_{1,t}$, $ \boldsymbol{\epsilon}_{2,t}$, and  $\boldsymbol{\epsilon}_{3,t}$ are the orthogonalized counterparts of  $\boldsymbol{u}_{1,t}$, $ \boldsymbol{u}_{2,t}$, and  $\boldsymbol{u}_{3,t}$, respectively. Moreover, the block inverse matrix inside equation (\ref{blockinverse}) is
\begin{align}
    \begin{pmatrix}
  \boldsymbol{\Sigma}_{11} &  \boldsymbol{\Sigma}_{12} \\
    \boldsymbol{\Sigma}_{21} &  \boldsymbol{\Sigma}_{22}
  \end{pmatrix}^{-1}
  \equiv 
      \begin{pmatrix}
  \boldsymbol{\Sigma}^{11} &  \boldsymbol{\Sigma}^{12} \\
    \boldsymbol{\Sigma}^{21} &  \boldsymbol{\Sigma}^{22}
  \end{pmatrix}
  =
   \begin{pmatrix}
  \boldsymbol{\Sigma}_{11}^{-1}+  \boldsymbol{\Sigma}_{11}^{-1} \boldsymbol{\Sigma}_{12}K \boldsymbol{\Sigma}_{21} \boldsymbol{\Sigma}_{11}^{-1} &-  \boldsymbol{\Sigma}_{11}^{-1} \boldsymbol{\Sigma}_{12}K\\
   K \boldsymbol{\Sigma}_{21} \boldsymbol{\Sigma}_{11}^{-1} & K
  \end{pmatrix}, \nonumber
\end{align}
  where 
  $$K=(\boldsymbol{\Sigma}_{22}-\boldsymbol{\Sigma}_{21}\boldsymbol{\Sigma}_{11}^{-1}\boldsymbol{\Sigma}_{12})^{-1}$$
   is the Schur complement of $\boldsymbol{\Sigma}$. Stacking equations (\ref{ortho1})-(\ref{blockinverse}) then yields
\begin{align} 
  \begin{pmatrix}
   \boldsymbol{\epsilon}_{1,t} \\
   \boldsymbol{\epsilon}_{2,t} \\
   \boldsymbol{\epsilon}_{3,t} \\
  \end{pmatrix}
  = & \nonumber
    \begin{pmatrix}
 \boldsymbol{u}_{1,t} \\
 \boldsymbol{u}_{2,t} \\
 \boldsymbol{u}_{3,t} \\
  \end{pmatrix} -
      \begin{pmatrix}
0 &0 &0\\
\boldsymbol{\Sigma}_{21}\boldsymbol{\Sigma}_{11}^{-1}&0 &0 \\
 \boldsymbol{\Sigma}_{31}\boldsymbol{\Sigma}^{11}+\boldsymbol{\Sigma}_{32}\boldsymbol{\Sigma}^{21}&\boldsymbol{\Sigma}_{31}\boldsymbol{\Sigma}^{12}+\boldsymbol{\Sigma}_{32}\boldsymbol{\Sigma}^{22}&0
  \end{pmatrix}
      \begin{pmatrix}
 \boldsymbol{u}_{1,t} \\
 \boldsymbol{u}_{2,t} \\
 \boldsymbol{u}_{3,t}
  \end{pmatrix}  \\
   = &  \nonumber
    \begin{pmatrix}
 \boldsymbol{I} &0 &0\\
-\boldsymbol{\Sigma}_{21}\boldsymbol{\Sigma}_{11}^{-1}& \boldsymbol{I}  &0 \\
- \boldsymbol{\Sigma}_{31}\boldsymbol{\Sigma}^{11}-\boldsymbol{\Sigma}_{32}\boldsymbol{\Sigma}^{21}&-\boldsymbol{\Sigma}_{31}\boldsymbol{\Sigma}^{12}-\boldsymbol{\Sigma}_{32}\boldsymbol{\Sigma}^{22}& \boldsymbol{I}
  \end{pmatrix}
      \begin{pmatrix}
 \boldsymbol{u}_{1,t} \\
 \boldsymbol{u}_{2,t} \\
 \boldsymbol{u}_{3,t}
  \end{pmatrix},  
  \end{align}
or
\begin{equation*} 
	  \boldsymbol{\epsilon}_t   =\boldsymbol{Q}^{-1}_3 \boldsymbol{u}_t.  
\end{equation*}

\addcontentsline{toc}{section}{References}
\bibliographystyle{Diebold}  
\bibliography{References}

  \end{document}